# Transparent perovskite barium stannate with high electron mobility and thermal stability

Woong-Jhae Lee,[1] Hyung Joon Kim,[1] Jeonghun Kang,[1] Dong Hyun Jang,[1] Tai Hoon Kim,[1] Jeong Hyuk Lee,[1] and Kee Hoon Kim[1,2]

[1]*Center for Novel States of Complex Materials Research, Department of Physics and Astronomy, Seoul National University, Seoul 151-747, Republic of Korea;*

[2]*Institute of Applied Physics, Department of Physics and Astronomy, Seoul National University, Seoul 151-747, Republic of Korea*

Email : khkim@phya.snu.ac.kr

**Abstract**

Transparent conducting oxides (TCOs) and transparent oxide semiconductors (TOSs) have become necessary materials for a variety of applications in the information and energy technologies, ranging from transparent electrodes to active electronics components. Perovskite barium stannate ($BaSnO_3$), a new TCO or TOS system, is a potential platform for realizing optoelectronic devices and observing novel electronic quantum states due to its high electron mobility, excellent thermal stability, high transparency, structural versatility, and flexible doping controllability at room temperature. This article reviews recent progress in the doped $BaSnO_3$ system, discussing the wide physical properties, electron-scattering mechanism, and demonstration of key semiconducting devices such as *pn* diodes and field-effect transistors. Moreover, we discuss the pathways to achieving two-dimensional electron gases at the interface between $BaSnO_3$ and other perovskite oxides and describe remaining challenges for observing novel quantum phenomena at the heterointerface.

# 1. INTRODUCTION

The term transparent conducting oxides (TCOs) often refer to materials having both high electrical conductivity ($\geq \sim 10^3$ S cm$^{-1}$) and high optical transmission ($\geq 80\%$) while the related term transparent oxide semiconductors (TOSs) represent materials having intermediate conductivity ($\sim 10^{-8}$ - $10^3$ S cm$^{-1}$) and high optical transmission ($\geq 80\%$). The combination of these two important features, i.e., makes TCOs or TOSs key materials for the passive or active components of various modern optoelectronic and electronic devices (1-5). For example, transparent electrical leads, *pn* junctions, and field-effect transistors (FETs) have been heavily used in liquid crystal displays (6), organic light-emitting diodes (7), solar cells (8), light-emitting diodes (9), and gas sensors (10), among other applications (11, 12) (**Figure 1***a*). Moreover, there exist increasing demands for the new TCOs or TOSs for next-generation devices related to the information and energy technologies.

Most known TOSs are *n*-type in which oxygen vacancies, ionized donor substitutions, and cation interstitials usually donate electrons to the conduction band, providing charge carriers for the flow of electric current (13). In addition, the most archetypal *n*-type TCO is Sn-doped In$_2$O$_3$ (ITO), which has been heavily used as passive transparent electrical leads by industries and universities for more than six decades (14). Other *n*-type TCOs include F-doped SnO$_2$ (15) and Al-doped ZnO (16); additional examples are given in References 17-32. These well-known *n*-type TCOs typically have electron carrier concentration (*n*) from $10^{18}$ to $10^{21}$ cm$^{-3}$ and an electron mobility (*μ*) from 10 to 100 cm$^2$ V$^{-1}$ s$^{-1}$, depending on material qualities and growth conditions. These rather high mobility values are due to the characteristic electronic structure of *n*-type TCOs, in which the conduction band is often composed of spatially delocalized, metallic *s* orbitals, resulting in high dispersion and low electron effective mass. Oxide materials with cations having a closed-shell configuration of $4s^0$ or $5s^0$—In(III), Sn(IV), Zn(II), Cd(II), and Ga(III)— show this characteristic condition (33, 34). For example, **Figure 1***b* shows the electronic band structure for In$_2$O$_3$ with a dispersive conduction band composed mainly of In 5*s* states (35). High dispersion and low effective mass are thus necessary for good electrical mobility in a given material system.

Because many known TCOs and TOSs have limitations, investigators are actively searching for alternative materials that can potentially exhibit better physical properties.

First, the overwhelming demand for ITO for use in transparent electrodes, coupled with the low natural abundance of In, has made In an increasingly expensive commodity, which has led to the search for an alternative material (36). Second, for the long-term operation of oxide electronic devices in air, the stability of TCOs is crucial, particularly at interfaces, to avoid the degradation problems frequently observed in oxide *pn* junctions and transistors. In many cases, degradation results from irreversible changes in oxygen defect structure, which can be exacerbated by thermal treatment. For example, the resistance of ZnO films at room temperature increased by $10^4$ times after heat treatment at 400°C for 2 h (37). Such unstable electrical properties stem from the thermal instability of oxygen. Instability of the oxygen defect structure can also cause various other phenomena such as (*a*) persistent current in the channel after photon irradiation and (*b*) unstable *p*-type properties (38, 39). Thus, the realization of stable *p*-type TCOs remains a major challenge in oxide electronics (40). Third, high carrier mobility is an important aspect for developing fast logic devices because carrier mobility fundamentally limits the operation speed across such devices (41, 42). Finally, controlling and minimizing the density of defects or dislocations are also important for achieving highly efficient devices based on TCOs and TOSs.

Materials with a perovskite structure have exhibited a plethora of interesting physical properties, such as large photovoltaic effects (43, 44), superconductivity (45-47), colossal magnetoresistance (48), ferroelectricity (49, 50), and multiferroicity (51, 52). Extensive research has been performed to utilize such versatile physical properties in the form of single-layered thin films and their heterostructures (53, 54). There is also growing interest in new perovskite TCOs, in which carrier doping and structural modification can be more flexible than in conventional TCOs made of binary oxides. The most-studied perovskite oxides in this context are doped SrTiO$_3$ and KTaO$_3$, in which electron mobility $\mu$ values at ~ 2 K are as high as 32,667 and 23,000 cm$^2$ V$^{-1}$ s$^{-1}$, respectively. (29, 55). These high $\mu$ are in part due to high dielectric constants ($\varepsilon_{STO}$ = ~20,000 and $\varepsilon_{KTO}$ = ~4,500) at low temperatures, which helps reduce the ionic dopant scattering due to high dielectric screening. However, at room temperature, those perovskite oxides generally showed much lower $\mu$ (~1–30 cm$^2$ V$^{-1}$ s$^{-1}$) (55, 56) than did those doped binary oxides, which has hindered practical device applications.

Alkaline-earth stannates with the chemical formula $A$SnO$_3$ ($A$ = Ca, Sr, and Ba) are other perovskite oxides that have recently gained attention. With systematic increase of the $A$-site effective ionic radius ($r_A$) from Ca$^{2+}$ (100 pm) to Sr$^{2+}$ (118 pm) to Ba$^{2+}$ (135 pm) (57), the structure changes from a distorted orthorhombic perovskite structure with space group *Pnma* (CaSnO$_3$ and SrSnO$_3$) to an ideal cubic perovskite with space group $Pm\bar{3}m$ (BaSnO$_3$) (58) (**Figure 1c**). At the same time, the Goldschmidt tolerance factor, defined by $(r_A + r_O)/\sqrt{2}(r_B + r_O)$, systemically increases from 0.927 to 0.961 to 1.018, with a corresponding reduction of the optical band gap ($E_g$) from 4.4 eV to 4.0 eV to 3.0 eV (59, 60). Although many novel physical properties await elucidation in this series of $A$SnO$_3$, we focus on BaSnO$_3$ (BSO) and its doped derivatives in this review.

Like all perovskites which crystallize in the ideal cubic $Pm\bar{3}m$ space group, the structure of BaSnO$_3$ is characterized as a three-dimensional framework of corner-sharing SnO$_6$ octahedra, in which the globally averaged value of the Sn-O-Sn bonding angle is 180° (**Figure 1c**). This bonding feature allows for enhanced electron hopping between neighboring Sn sites, constituting a physical origin for conduction bands with high dispersion. **Figure 1b** presents the electronic band structure of BaSnO$_3$ as calculated by density functional theory (DFT) calculations within the local density approximation (LDA) formalism. The conduction band, composed mainly of the Sn 5$s$ state, is quite dispersive (i.e., large bandwidth) (61). Similar band calculation results can also be found in the earlier works of Singh et al. (62) in 1991 and Mizoguchi et al. (60) in 2004. Such high dispersion and wide band gap are common properties of both BaSnO$_3$ and In$_2$O$_3$, although these materials have completely different crystal structures and electronic states in the conduction band. Accordingly, Mizoguchi et al. (60) first predicted that BaSnO$_3$ with high dispersion could potentially be a good TCO. However, the measured $\mu$ by Wang et al. (63), Hadjarb et al. (64), and Liu et al. (65) in the thin films and polycrystals of electron-doped BaSnO$_3$ was relatively low (< 1 cm$^2$ V$^{-1}$ s$^{-1}$), and the maximum conductivity was 250 S cm$^{-1}$. Those early specimens were likely subject to severe extra-scattering, possibly from poor stoichiometry control, poor crystallinity, or the presence of grain boundaries. Therefore, there has been high demand for the growth of high-quality single crystals of doped BaSnO$_3$ to understand their intrinsic physical properties.

With the motivations of searching for an alternative TCO and uncovering the intrinsic physical properties of doped BaSnO$_3$, we have systematically grown single crystals and high-quality thin films of (Ba,La)SnO$_3$ (BLSO) and Ba(Sn,Sb)O$_3$ (BSSO) since 2009. With these efforts, we have discovered that BaSnO$_3$ single crystals with few-percent La doping produced unusually high $\mu$ (up to 320 cm$^2$ V$^{-1}$ s$^{-1}$ at room temperature) and optical transparency with enhanced $E_g$ ($\geq$ 3.1 eV) (32, 61). This value of $\mu$ = 320 cm$^2$ V$^{-1}$ s$^{-1}$ at room temperature was the highest among the popular TCOs in a degenerate semiconductor regime. Moreover, we found superior thermal stability in the electrical properties of doped BaSnO$_3$, i.e., a less-than 2% change in conductivity in air atmosphere at a high temperature of 530°C (32). Therefore, the doped BaSnO$_3$ system could potentially overcome the problems caused by changes in the oxygen defect structure in oxide electronic devices (66). More importantly, the new perovskite stannates with both high $\mu$ and stability could be combined with other perovskite materials with diverse physical properties and could thus be key to realizing all-perovskite oxide electronic devices having the exotic physical properties of constituent materials or having unexpected novel quantum phenomena at the heterointerfaces.

In this review, we summarize research since 2012 and the potential opportunities of science and applications related to the doped BaSnO$_3$ system. As this is the first review on this exciting transparent conducting system, we also provide a detailed outlook for future research.

## 2. DISCOVERY OF BASNO$_3$–BASED SINGLE CRYSTALS

### 2.1. High-Quality Single-Crystal Growth and High Electron Mobility

Because the electron mobility $\mu$ in a simple one-band model is defined as

$$\mu = e\tau / m^*, \quad 1.$$

where $\tau^{-1}$ and $m^*$ are the total scattering rate and the electron effective mass, respectively, one might expect high $\mu$ once lower $m^*$ or $\tau^{-1}$ is realized. The low $m^*$ of a semiconductor is a natural consequence of the dispersive conduction band, particularly near the band minimum. Indeed, the conduction band of the representative TCOs is quite dispersive and has rather low $m^*$ of 0.1 – 0.5$m_0$, where $m_0$ is the electron mass. Motivated by the dispersive conduction band of BaSnO$_3$ (**Figure 1b**), our group first tried to grow single crystals of electron-doped BaSnO$_3$ (32). In 2010 and 2011, we successfully grew BaSnO$_3$ (69), BLSO (61), and BSSO (69) single crystals by the flux method in air. The flux method has been frequently used for single-crystal growth when the seed materials for forming the target material can be dissolved in a flux (solvent). For the flux, we tried to use PbF$_2$. We mixed the flux with polycrystalline BaSnO$_3$ powder at a ratio BaSnO$_3$:PbF$_2$ = 1:2.5 and fired at 1,200°C for 5 h, followed by a slow cool down to 900°C at a rate of 5°C h$^{-1}$. We obtained relatively large BaSnO$_3$ single crystals with a reddish color. This initial attempt made us conclude that Pb impurities can enter Ba or Sn sites, which can increase carrier scattering in the doped system. We then tried other fluxes for a significant amount of time and eventually found that a mixture of Cu$_2$O and CuO powders with a molar ratio of Cu$_2$O:CuO = 61:39 was an excellent flux for our purpose. At this molar ratio, the Cu$_2$O + CuO flux formed an eutectic melting point at 1,090°C (67), which is lower than the melting temperature of Cu$_2$O flux (1,230°C) by 140°C. For growth, we put a mixture of the BaSnO$_3$ powder and the flux into the Pt crucible with various molar ratios of BaSnO$_3$:flux materials = 1:1–100. The mixture was fired for 5 h at 1,250°C and was then slowly cooled to 1,070°C at a 1°C h$^{-1}$ rate.

We obtained BLSO single crystals with a typical size of ~2 × 2 × 2 mm$^3$ and a maximum size of 5 × 3 × 1 mm$^3$. Most importantly, using electron probe microanalysis (EPMA), we could not observe any Cu-related impurities in the grown BaSnO$_3$, BLSO, or BSSO crystals (61). Therefore, any impurity levels resulting from the Cu$_2$O and CuO flux are much lower than the resolution of EPMA. We thus expected that those single crystals, without any Cu impurities from the flux, could offer the possibility of observing the intrinsic properties (i.e., intrinsic mobility) of BaSnO$_3$, BLSO, and BSSO crystals. Intrinsic mobility is expected to show up once the extra-scattering sources, e.g., ionized and neutral impurities, are minimized. Luo et al. (68) also reported that single crystals grown by PbO-based flux produced a lower $\mu$ ~103 cm$^2$ V$^{-1}$ s$^{-1}$ with $n$ = 1.0 × 10$^{20}$ cm$^{-3}$. This observation of lower $\mu$ implies that Pb impurities can cause extra-scattering.

On the basis of the Cu$_2$O + CuO flux, we successfully grew not only high-quality BaSnO$_3$ but also BLSO single crystals exhibiting high $\mu$. **Figure 2** compares the $\mu$ of BLSO single crystals at room temperature with the $\mu$ values of several representative semiconductors and TCOs. Although located in the second-highest position after GaAs, BLSO exhibits the highest $\mu$ of 320 cm$^2$ V$^{-1}$ s$^{-1}$ among representative TCOs and TOSs in a degenerate *n*-type doped regime ($n \geq 10^{19}$ cm$^{-3}$) (21, 56, 61, 69-74). It shows even higher $\mu$ than GaN, which is the key semiconductor for blue light–emitting diodes (75). Moreover, at room temperature, the $\mu$ of BLSO is 30 times higher than that of doped SrTiO$_3$ (56, 71), the most popular semiconducting perovskite oxide.

In comparison with Ba-site doping, we also studied the effect of Sn-site doping with Sb to form BSSO. At least half of the Sb dopants were activated at room temperature to behave as Sb$^{5+}$ and to provide extra electron carriers (69). As summarized in **Figure 2**, the $\mu$ of BSSO single crystals reached ~79 cm$^2$ V$^{-1}$ s$^{-1}$ at $n = 1.0 \times 10^{20}$ cm$^{-3}$, and upon further increases in $n$, $\mu$ systematically decreased, being nearly proportional to $n^{-1}$, which supports dominant neutral impurity scattering. According to our estimation of electron carriers with nominal Sb doping, half of the Sb ions remained as neutral impurities to give rise to additional scattering (see Section 5.3). Our results indicate that the *B*-site dopant may cause more scattering than the *A*-site dopant, as the Sb ions are located in the direct conduction path of Sn-O-Sn bonding.

Meanwhile, nominally undoped BaSnO$_3$ crystals show *n*-type carriers, which are presumably caused by the oxygen vacancies created during crystal growth at high temperatures under air (69). The $\mu$ of oxygen-deficient BaSnO$_3$ (BaSnO$_{3-\delta}$) single crystals was 180 cm$^2$ V$^{-1}$ s$^{-1}$ at $n = 3.0 \times 10^{18}$ cm$^{-3}$ (**Figure 2**), suggesting that the number density of oxygen vacancies is ~$1.5 \times 10^{18}$ cm$^{-3}$ because each vacancy yields two electron carriers. The grown crystal could be used as a metallic substrate (electrical conductivity $\sigma$ at room temperature = ~100 S cm$^{-1}$), similar to the case of Nb:SrTiO$_3$. However, for general use as highly insulating substrates, it was necessary to find a new growth method.

## 2.2. Growth of Highly Insulating BaSnO$_3$ Single Crystals

On the basis of the observation of dislocation-limited transport in BLSO/SrTiO$_3$(001) films, which is discussed in Section 3.1 below, reducing dislocation densities is important. The use of a substrate with cubic lattice constant ($a_c$) ~4.116 Å is obviously a promising direction. However, there are no commercial substrates whose $a_c$ is close to ~4.116 Å. Moreover, due to highly evaporating and decomposing nature of constituent elements, the growth of large BaSnO$_3$ single crystals has been challenging. Second, the best candidate substrate, BaSnO$_3$, cannot be grown bigger than ~5 × 5 mm$^2$ within the flux method. The other growth methods that can potentially result in bigger sizes, such as the Czochralski and optical floating-zone methods, have so far failed due to high evaporation and decomposition of constituent elements. Third, BaSnO$_3$ single crystals grown by the flux method still contain impurities from fluxes such as Pb ions (68) and oxygen vacancies (69). Before our work in 2016 (81), there were no insulating BaSnO$_3$(001) single crystals as a substrate. Finding a suitable substrate is therefore one of the main hurdles to overcome in order to grow high-quality thin films.

An ideal substrate for doped BaSnO$_3$ films would be insulating BaSnO$_3$ single crystals, as the $a_c$ = 4.116 Å of these single crystals matches well with the $a_c$ of BLSO within 0.1%. However, there are no commercial BaSnO$_3$ substrates so far, and the growth of highly insulating and large BaSnO$_3$ single crystals has been challenging. Our first attempt to grow BaSnO$_3$ crystals was by the flux method, using a KF flux. This resulted in small BaSnO$_3$ crystals with a diameter of less than 0.1 mm. Larger single crystals up to ~10 mm$^2$ were grown by PbF$_2$-based flux, but they contained significant Pb impurities and consequently had reduced mobility and crystallinity (68). With Cu$_2$O + CuO flux, single crystals of typically ~1 – 8 mm$^2$ could be grown without significant Cu impurities. However, these crystals usually contained oxygen vacancies, which resulted in *n*-type carriers at a concentration 10$^{19}$ cm$^{-3}$ (69).

In accordance with the requirement of an ideal substrate, we have tried to grow large-area insulating BaSnO$_3$ single crystals by use of a Pt crucible with a volume of ~400 mL and by increasing the temperature gradient during growth. We succeeded in synthesizing highly insulating BaSnO$_3$ single crystals by adding KClO$_4$ to the Cu$_2$O-CuO flux (**Supplemental Figure 1*a,b***). The maximum size of insulating BaSnO$_3$ single crystal is ~3 × 3 × 3 mm$^3$ (**Figure 3***a*). The molar ratio that resulted in the most insulating properties ($\rho$ = 10$^{12}$ Ω·cm at 300 K) was BaSnO$_3$:(Cu$_2$O + CuO):KClO$_4$ = 1:50:2 (**Supplemental Figure 1**). The EPMA study on those BaSnO$_3$ single crystals revealed that they contained a detectable amount of K (~1.4 × 10$^{19}$ cm$^{-3}$), indicating that the *n* were partially compensated by hole carriers from K impurities (81). In comparison with previously grown BaSnO$_{3-\delta}$ showing metallic resistivity of 1 mΩ·cm, BaSnO$_3$ single crystals possessed superior insulating characteristics. Moreover, the rocking curve of the (002) Bragg peak produced a FWHM (full width at half-maximum) as small as 0.02°, demonstrating superior crystallinity (81).

### 2.3. Toward New Substrates with Better Lattice-Matching Properties

Researchers are currently attempting to grow thin films by using substrates with an $a_c$ or the pseudocubic lattice constant ($a_{pc}$) (76) close to that of BaSnO$_3$. In **Figure 3d**, we summarize the band gap versus $a_c$ (or $a_{pc}$) of the cubic (or orthorhombic) perovskite materials that have been used or can be used potentially as substrates for growing BaSnO$_3$ films. **Supplemental Table 1** includes detailed data and references for **Figure 3d**.

In addition to these known substrates, there are recent research efforts to find new substrate materials. Uecker et al. (114) recently synthesized large single crystals of (LaLuO$_3$)$_{1-x}$(LaScO$_3$)$_x$ with $a_{pc}$ between 4.09 and 4.18 Å by the Czochralski method. The grown crystals had a length of approximately 75 mm and a diameter of 15 mm (**Figure 3b**). We anticipate that new single crystals of (LaLuO$_3$)$_{1-x}$(LaScO$_3$)$_x$ for which $a_c$ is near 4.116 Å could be useful for growing high-quality BaSnO$_3$ films. In another effort, our group recently succeeded in growing large LaInO$_3$ single crystals by the optical floating-zone method (**Figure 3c**). LaInO$_3$ forms an orthorhombic perovskite structure with the *Pnma* space group, and the cell parameters are $a = 5.9414$ Å, $b = 8.2192$ Å, and $c = 5.7249$ Å. Although LaInO$_3$ has an orthorhombic distortion, its $a_{pc} = 4.119$ Å is close to that of BaSnO$_3$. The length of the grown crystal is ~50 mm, and its diameter is ~7 mm. The main direct gap and $\varepsilon_r$ are 4.13 eV and 23.7, respectively, demonstrating that the system is another transparent material with a rather high $\varepsilon_r$. Indeed, several theoretical and experimental studies show that LaInO$_3$ is a polar material at the atomic scale [due to sequential layers of (LaO)$^+$ and (InO$_2$)$^-$] and can thus induce two-dimensional electron gases (2DEGs) and electrical conductivity in a LaInO$_3$-BaSnO$_3$ heterostructure (97, 116). Therefore, systematic studies of uniaxial or biaxial strain effects on the physical properties of LaInO$_3$ become necessary for designing and understanding possible 2DEG behavior at the LaInO$_3$-BaSnO$_3$ heterointerface.

### 3. IMPROVED MOBILITY IN DONOR-DOPED BASNO$_3$ FILMS

In this section, we summarize recent research progress in improving $\mu$ in electron-doped BaSnO₃ films. For convenience, we divide those into three categories. The first category encompasses the result of BLSO films on SrTiO₃(001) substrates grown by the pulsed laser deposition (PLD) technique. The second category encompasses research using substrates with lower lattice misfit or buffer layers. The third category encompasses research relying on the molecular beam epitaxy (MBE) method, which generally results in better film quality relative to the other growth techniques.

### 3.1. Thin Films of (Ba,La)SnO₃ on SrTiO₃(001) substrates

One necessary step for future applications of the BaSnO₃ system is growing high-quality thin films with high $\mu$. In 2012, we found that epitaxial BLSO films grown on SrTiO₃(001) substrates [BLSO/SrTiO₃(001)] showed a $\mu$ of ~70 cm² V⁻¹ s⁻¹ at $n = 4.4 \times 10^{20}$ cm⁻³ (**Figure 2**) and that the FWHM of the rocking curve of the BLSO(002) Bragg peak was 0.09° (32, 61). In the earlier reports of epitaxial BLSO films grown on SrTiO₃(001) or MgO(001), the FWHM in the rocking curve was 0.14 - 0.57° (63-65). Thus, the transport properties of even epitaxial thin films should be sensitive to the thin films' crystallinity and local microstructures. Therefore, it is worthwhile to understand the close relationship between the electrical properties and microstructural properties of such thin films.

Until recently, most reported BLSO films were grown on SrTiO₃(001) substrates, whose $a_c = 3.905$ Å has a large misfit of 5.13% relative to the lattice constant of BaSnO₃(001) ($a_c = 4.116$ Å). Panels *a* and *b* of **Figure 4** show low- and high-magnification cross-sectional transmission electron microscopy (TEM) images, respectively, at the film-substrate interface of Ba₀.₉₉₅La₀.₀₀₅SnO₃/SrTiO₃(001). Misfit dislocation arrays in the BLSO film with a period = ~7.4 - 7.8 nm were found (66). These values correspond to the 19 - 20 layers of {101} planes in the SrTiO₃ substrate and to the 18 - 19 layers of the {101} planes in the BLSO film. Therefore, the main source for the dislocations should be the SrTiO₃(001) substrate, which has a large lattice mismatch with BaSnO₃.

Naturally, the best $\mu$ of BLSO/SrTiO₃(001) was limited to 40 - 70 cm² V⁻¹ s⁻¹ at high $n$ (> 2 × 10²⁰ cm⁻³) and was lower than 30 cm² V⁻¹ s⁻¹ in a low-$n$ regime (< 7 × 10¹⁹ cm⁻³) (**Figure 2**). The characteristic decrease of $\mu$ with $n$, which is roughly proportional to ~$n^{1/2}$ (112, 113), supports dominant dislocation scattering. This is in sharp contrast to the behavior of single crystals exhibiting a nearly $n$-independent $\mu$ of 200–300 cm$^2$ V$^{-1}$ s$^{-1}$ due to dominant ionized dopant scattering (see Section 5.2). Therefore, the $\mu$ in the thin films can increase by using a substrate with a smaller misfit to reduce crystalline defects.

### 3.2. (Ba,La)SnO₃ Thin Films with Enhanced Mobility

Several studies have improved crystallinity and $\mu$ of films by using substrates with reduced lattice misfits. Substrates with an orthorhombic structure include SmScO$_3$(110)$_O$, TbScO$_3$(110)$_O$, and PrScO$_3$(110)$_O$. Here, the subscripted O refers to the orthorhombic index. For these structures, the pseudocubic lattice constant $a_{pc}$ (76) and lattice misfit compared with BaSnO$_3$ are as follows: SmScO$_3$(110)$_O$ (3.868 Å, −3.07%), TbScO$_3$(110)$_O$ (3.958 Å, −3.84%), and PrScO$_3$(110)$_O$ (4.021 Å, −2.30%) (**Figure 3d**). **Figure 5** shows $\mu$ at room temperature as a function of $n$ for electron-doped BaSnO$_3$. In 2014, Wadekar et al. (77) fabricated BLSO films on SmScO$_3$(110)$_O$ substrates by the PLD technique and found $\mu$ = ∼10 cm$^2$ V$^{-1}$ s$^{-1}$. As an alternative idea to enhance $\mu$, Park et al. (79) deposited a BaSnO$_3$ buffer (thickness $t$ = 110 nm) on SrTiO$_3$(001) substrate before growing BLSO films to find reduced threading dislocations and enhanced $\mu$ = ∼80 cm$^2$ V$^{-1}$ s$^{-1}$ in the BLSO film. Shiogai et al. (80) also grew BLSO films grown on (Ba,Sr)SnO$_3$ buffer ($t$ = ∼200 nm)/SrTiO$_3$(001), finding $\mu$ = 78 cm$^2$ V$^{-1}$ s$^{-1}$ at $n$ = 1.6 × 10$^{20}$ cm$^{-3}$.

By using an insulating BaSnO$_3$(001) single crystal as a substrate, we could grow successfully epitaxial BLSO films on highly insulating BaSnO$_3$(001) substrates [BLSO/BaSnO$_3$(001)] by the PLD method (81). The $\mu$ at room temperature was as high as 102 cm$^2$ V$^{-1}$ s$^{-1}$ at $n$ = 1.0 × 10$^{20}$ cm$^{-3}$ and slightly increased with an decrease in $n$, being qualitatively similar to single-crystal behavior (**Figure 5**). This is currently the highest $\mu$ value among BLSO films grown by PLD. Moreover, we could not detect any dislocations or grain boundaries in a bright-field TEM image that covers a wide area of the film cross section over ∼300 × 100 nm$^2$ (**Figure 4c,d**). This high-resolution TEM image also demonstrated that the lattice periodicity of the BLSO film almost perfectly matches that of the BaSnO$_3$ substrate. However, as $\mu$ = 102 cm$^2$ V$^{-1}$ s$^{-1}$ at room temperature is still lower than the highest value of 320 cm$^2$ V$^{-1}$ s$^{-1}$ of the single crystals, other scattering sources should exist in thin films. Point defects due to cation vacancies or cation site mixing may exist in BLSO/BaSnO$_3$(001) films, as indicated by the TEM image exhibiting broad dark and white contrasts in **Figure 4c**. Similar kinds of point defects exist in SrTiO$_3$ films on a SrTiO$_3$(001) substrates grown by the PLD method (117). Thus, there is much room for improvement with regard to $\mu$. Efforts to increase $\mu$ in BaSnO$_3$-based films by use of better substrates currently constitute an active research area.

### 3.3. Growth by the MBE Method

To achieve 2DEGs at the interfaces of BaSnO$_3$-related heterostructures, defect-free BaSnO$_3$ films should be a prerequisite so that film growth in a layer-by-layer mode becomes necessary. Until 2015, most film growth for the BaSnO$_3$ system had been limited to PLD (61, 77-81, 101, 118-121), sputtering (122), and solution deposition (123), resulting in relatively low $\mu$, as summarized in **Figure 5**. The $\mu$ values of BLSO films deposited by those growth methods are between 10 and 102 cm$^2$ V$^{-1}$ s$^{-1}$ at $n$ ranging from $5 \times 10^{19}$ to $8 \times 10^{20}$ cm$^{-3}$. Even though epitaxial BLSO films could be grown by the PLD technique, they still had considerable point defects, which formed due to the high-energy particles involved during the growth process (81, 119).

Prakash et al. (124) first reported, in 2015, oxide MBE growth of BaSnO$_3$, which can potentially overcome these defect problems. They grew epitaxial phase-pure, stoichiometric BaSnO$_3$ films on SrTiO$_3$(001) and LaAlO$_3$(001) substrates. They reported intensity oscillations of reflection high-energy electron diffraction (RHEED), a characteristic feature of the layer-by-layer growth mode, during growth. In 2016, Lebens-Higgins et al. (78) fabricated BLSO films on TbScO$_3$(110)$_O$ substrates by the oxide MBE method to find $\mu$ = 81 cm$^2$ V$^{-1}$ s$^{-1}$. The grown films showed superior crystallinity, with a FWHM of 0.006° in the rocking curve of the BLSO(002) peak, which is a record low value. Subsequently, Raghavan et al. (31) reported other MBE-grown BLSO films on PrScO$_3$(110)$_O$ substrates with room-temperature $\mu$ = 150 cm$^2$ V$^{-1}$ s$^{-1}$ by using preoxidized Sn as a precursor. It is noteworthy that this is the highest-known $\mu$ in doped BaSnO$_3$ films. Those different mobility values imply that electrical properties depend on the precursors and the substrates. Therefore, there seem to be scope for improvement in $\mu$ in thin stannate films grown by the MBE technique. Due to MBE's advantage with regard to controlling defects, this technique is expected to allow the growth of thin stannate films with higher $\mu$ in a broad doping regime, particularly in a low-dopant regime.

Due to the above-described progress, there is great hope for fabricating higher-quality films based on BaSnO$_3$(001) substrates by using, e.g., the oxide MBE or laser MBE techniques, which can reduce the point defects resulting from cation mixing or cation deficiencies. Yet the current maximum size of insulating BaSnO$_3$ substrates is only approximately 3 × 3 mm$^2$, and thus thin-film growth is still difficult. One future challenge is to grow bigger BaSnO$_3$ single crystals with highly insulating properties.

## 4. THERMAL STABILITY AND OTHER PHYSICAL PROPERTIES

### 4.1. Superior Thermal Stability of Oxygen

Thermal instability, associated with the behavior of lattice oxygen, is among the most intriguing of the physical properties of oxide thin films, and has become a major bottleneck in the device applications. Many electronic oxides are often subject to severe degradation due to such instability. The example of ZnO has already been described in Section 1. Moreover, in perovskite titanates, whether there is oxygen deficiency at the interface remains controversial in spite of much ongoing research on their 2DEG behavior (82, 83). Furthermore, in oxide-based resistive memory devices, the thermal stability of oxygen at the interface and/or the reaction of oxide with hydrogen and water fundamentally govern the electrical transport behavior of these devices under electric field (84, 85).

In general, as known in doped perovskite oxides (e.g., titanates), at high temperatures more than 1000ºC, cation vacancies are predominantly created at high oxygen partial pressures to compensate donor-type dopants (150, 151) whereas in doped BaSnO$_3$ at relatively low temperatures ~500ºC, the reduction of oxygen vacancies is dominant over the creation of cation vacancies (66). Therefore, at temperatures close to 500ºC, the conductivity of doped BaSnO$_3$ can be mainly controlled by oxygen vacancies in both reducing and oxidizing atmosphere. We have therefore systematically investigated the extent to which the oxygen stoichiometry in the BLSO system is stable from the transport studies around 500ºC.

Panels *a* and *b* of **Figure 6** present the time- and temperature-dependent resistance (*R*) measurements of a BLSO/SrTiO$_3$(001) film under Ar, O$_2$, and air atmosphere (32). We also measured the time-dependent Hall effect upon a change of gas atmosphere at a fixed temperature of 530°C (66) (**Figure 6c**). In the isothermal measurements, significant variations of $R$, $n$, and $\mu$ were observed upon changing the gas atmosphere from e.g., Ar to $O_2$. The $R$ in particular could be described by two exponential variations. The first fast response was attributed to surface reaction kinetics of Ar or $O_2$, occurring on a time scale typically less than 2 h. The slow responses occurred in a time scale of more than a few hours and resulted in an $R$ change of only about approximately 8% (9.5%) under Ar ($O_2$) atmosphere. The slow response was assigned to oxygen diffusion out (in) of the film under Ar ($O_2$) atmosphere. In the final annealing in air (**Figure 6b**), the film showed a very small decrease in $R$, only 1.7%, when it stayed at 530°C for 5 h, demonstrating the film's superior thermal stability in air. We observed a change in $n$ corresponding to the observed change in $R$ suggesting the change in $n$ is a major factor of $R$ variation during the oxygen diffusion process in the BLSO films (32).

When the slow oxygen diffusion-in or -out phenomena at the film surface was approximately modelled as one-dimensional diffusion along the film thickness direction, the chemical diffusion coefficient of oxygen within the BLSO film layer could be extracted by fitting time-dependent total $n$ inside the film from Hall effect measurements, yielding chemical diffusion coefficient of oxygen as low as $10^{-16}$ cm$^2$ s$^{-1}$ at 530°C (66). This value is much lower than the typical chemical diffusion coefficient of oxygen in other perovskite oxides, e.g., $SrTiO_3$ [~$10^{-7}$ cm$^2$ s$^{-1}$ (86)], $BaCeO_3$ [~$10^{-9}$ cm$^2$ s$^{-1}$ (87)], $BaBiO_3$ [~$10^{-7}$ cm$^2$ s$^{-1}$ (88)], and $LaCoO_3$ [~$10^{-6}$ cm$^2$ s$^{-1}$ (88)]. This finding provides an explanation for why the transport properties of BLSO are rather stable under the change of gas atmosphere as compared with other perovskite oxides.

### 4.2. Other Physical Properties of BaSnO$_3$

In **Figure 7**, we summarize various physical properties of the electron-doped BaSnO$_3$ system from band structures (61, 69), as well as this system's optical and thermal properties (61, 69, 89-92) and electrical properties (61, 69, 81), for general readers who might be interested in those basic properties. We refer readers to the **Supplemental Material**.

## 5. ORIGIN OF HIGH ELECTRON MOBILITY

In this section, we discuss the origins for achieving such a high $\mu$ in BLSO systems based on the peculiar characteristics reflected in two physical parameters, i.e., electron effective mass and electron-scattering rate.

## 5.1. Electron Effective Mass of BaSnO$_3$

To understand how electron-doped BaSnO$_3$ systems achieve high mobility, one should expect a low $m^*$ in the $\Gamma$ point of the electronic band structure. $m^*$ values of BaSnO$_3$ and electron-doped BaSnO$_3$ vary from 0.03 to 0.6$m_0$, according to theoretical and experimental results. In theory, DFT calculations predicted somewhat similar but varying values for each exchange-correlation functional: $m^* = 0.028m_0$ (93) with an LDA functional; $m^* = 0.029m_0$ (93) with a generalized gradient approximation (GGA) functional; $m^* = 0.47m_0$ (94), 0.20$m_0$ (95), 0.21$m_0$ (96), and 0.26$m_0$ (97) with a Heyd-Scuseria-Ernzerhof (HSE06) hybrid functional; and $m^* = 0.22m_0$ (98) with a Perdew-Burke—Ernzerhof (PBE0) hybrid functional. In experiments, various $m^*$ values were obtained: $m^* = 0.31m_0$ (99) and 0.61$m_0$ (69) from optical transmission, $m^* = 0.35m_0$ (100) and $m^* = 0.27m_0$ (101) from infrared spectroscopy, $m^* = 0.42m_0$ (102) from ellipsometry, and $m^* = 0.14m_0$ (103) from magnetic field–dependent photoluminescence.

Except for one DFT calculation predicting ~0.03$m_0$ (93), most values are located in a range from 0.2 to 0.6$m_0$. They are lower than those of titanium-based perovskites—CaTiO$_3$ [4.0$m_0$ (104)], SrTiO$_3$ [4.8$m_0$ (104)], and BaTiO$_3$ [5.3$m_0$ (104)]—by roughly one order of magnitude, showing that BaSnO$_3$ can be a useful mother compound for developing TCOs and TOSs with a high $\mu$. Interestingly, these $m^*$ values of BaSnO$_3$ are not considerably lower than those of common TCOs and semiconductors, e.g., In$_2$O$_3$ [0.35$m_0$ (14)], SnO$_2$ [0.25$m_0$ (105)], ZnO [0.28$m_0$ (106)], Si [0.27$m_0$ (107)], and GaN [0.20$m_0$ (75)]. Even the lowest predicted value of $m^* = 0.2m_0$ for BaSnO$_3$ does not fully account for why $\mu$ of BaSnO$_3$ is higher than $\mu$ of other TCOs, GaN, and Si in a degenerate doping regime (**Figure 2**). This insignificant difference between the $m^*$ values of BaSnO$_3$ and those of other materials implies that the BaSnO$_3$ system should have an unusually low scattering rate.

## 5.2. Scattering Due to Phonons and Ionized Impurities in (Ba,La)SnO$_3$ and BaSnO$_{3-\delta}$

When multiple scattering sources exist, the scattering rates of individual sources ($\tau_i^{-1}$) should contribute to determining a total electron-scattering rate ($\tau^{-1}$). In the momentum-independent approximation as well as in a weak scattering limit, $\tau^{-1}$ can be simply calculated as a sum of several $\tau_i^{-1}$ according to Matthiessen's rule:

$$\tau^{-1} = \sum_i \tau_i^{-1}. \quad\quad 2.$$

Similarly, the total electron mobility ($\mu$) can be determined by the mobility for the *i*th scattering source $\mu_i$,

$$\mu^{-1} = \sum_i \mu_i^{-1}. \quad\quad 3.$$

In bulk materials with minimized extrinsic scattering sources, e.g., grain boundaries and crystallographic defects, there are two major scattering mechanisms: acoustic phonon scattering and ionized impurity scattering. Both can be important for scattering analyses in a degenerate semiconductor. However, as the phonon scattering can be minimized at very low temperatures, comparison of the $\mu$ of BLSO crystals at 300 K ($\mu_{300K}$ = ~300 cm$^2$ V$^{-1}$ s$^{-1}$) and at 2 K ($\mu_{2K}$ = ~600 cm$^2$ V$^{-1}$ s$^{-1}$) in **Figure 7e** predicts that the scattering rate due to acoustic phonons is roughly half the total scattering rate at room temperature. In other words, from the relationship

$$\frac{1}{\mu_{300K}} = \frac{1}{\mu_{2K}} + \frac{1}{\mu_{ph}}, \quad\quad 4.$$

one can estimate $1/\mu_{ph}$ = ~1/600 (V·s) cm$^{-2}$, yielding $\mu_{ph}$ = ~600 cm$^2$ V$^{-1}$ s$^{-1}$. This analysis suggests that when ionized dopant scattering in BLSO crystals could be eliminated, the $\mu$ of BLSO crystals could be as high as 600 cm$^2$ V$^{-1}$ s$^{-1}$ at room temperature. Similarly, the low temperature (~ 2 K) $\mu$ of BaSnO$_3$ single crystals with low $n \leq$ ~10$^{18}$ cm$^{-3}$ could exceed 1,000 cm$^2$ V$^{-1}$ s$^{-1}$ if ionized impurity scattering can be made negligible. This rough estimation makes observation of quantum phenomena at low temperatures of doped BaSnO$_3$ a highly promising prospect. In application of **Equation 4**, scattering sources other than phonons and ionized impurities were neglected (69). Therefore, it is important to fabricate high-quality specimens without extra scatterings from grain boundaries/dislocations and crystallographic defects.

In a degenerate semiconductor, the ionized impurity scattering is independent of temperature, whereas the phonon scattering is temperature dependent (108). Therefore, to obtain a better understanding of the transport properties of BLSO with a degenerate electron gas, $\mu$ as a function of $n$, instead of as a function of temperature, should be investigated. $\mu$ is then determined mostly by scattering by ionized impurities (intrinsic lattice defects or extrinsic dopants). These impurities are screened by carriers owing to the charged Coulomb potential, thereby resulting in $\mu$ being dependent on $n$. The expressions derived by Dingle (109) can be used to describe this situation:

$$\mu_{ii} = \frac{3(\varepsilon_r \varepsilon_0)^2 h^3}{Z^2 m^* e^3} \frac{n}{N_i} \left[ \ln(1+\xi) - \frac{\xi}{1+\xi} \right]^{-1}, \quad 5a.$$

$$\xi = (3\pi^2)^{1/3} \frac{\varepsilon_r \varepsilon_0 h^2 n^{1/3}}{m^* e^2}, \quad 5b.$$

where $h$ is Planck's constant, $\varepsilon_r$ is the relative dielectric permittivity, $N_i$ is the density of ionized impurities, and $Z$ is the effective charge of ionized impurities (i.e., $Z = +1$ for $La^{3+}$ and $Sb^{5+}$, and $Z = +2$ for oxygen vacancy). For BLSO single crystals, almost full activation of the La impurities ($n = N_i$) was observed from Hall effect measurements (69). Finally, the two mobility expressions in **Equation 4** and **Equation 5a**, i.e., $\mu_{ph}$ and $\mu_{ii}$, can be combined:

$$\frac{1}{\mu_{sc}} = \frac{1}{\mu_{ph}} + \frac{1}{\mu_{ii}}, \quad 6.$$

where $\mu_{sc}$ is the total $\mu$ of single crystals. For example, $\mu_{BLSO}$, $\mu_{BSSO}$, and $\mu_{BSO-\delta}$ refer to the total mobility of BLSO, BSSO, and $BaSnO_{3-\delta}$ single crystals, respectively. $\mu_{ii}$ is the mobility curve limited by ionized impurities; $\mu_{ii\_D}$ and $\mu_{ii\_ov}$ apply more specifically to the cases of La/Sb dopants and oxygen vacancies (ov), respectively. **Figure 8** shows the thus-calculated $(\mu_{ii\_D}^{-1} + \mu_{ph}^{-1})^{-1}$ and $(\mu_{ii\_ov}^{-1} + \mu_{ph}^{-1})^{-1}$ curves based on **Equation 6** by use of the main experimental parameters of $m^* = 0.4 m_0$ and $\varepsilon_r = 20$ (110). In this calculation, it is assumed that $\mu_{ph}$ (~600 cm$^2$ V$^{-1}$ s$^{-1}$) is independent of $n$. Then, the corresponding expressions for the ionized impurity scattering shown in **Equation 5a** and **Equation 5b** are used to calculate total predicted mobility ($\mu_{ii\_D}^{-1} + \mu_{ph}^{-1})^{-1}$ and ($\mu_{ii\_ov}^{-1} + \mu_{ph}^{-1})^{-1}$, as represented in **Figure 8**. The resultant lines roughly agree well with the experimental data of BLSO and BaSnO$_{3-\delta}$ crystals. In particular, the $\varepsilon_r = 20$ of BaSnO$_3$ is almost two times higher than the $\varepsilon_r$ of the well-known TCOs In$_2$O$_3$ ($\varepsilon_r = 9$), SnO$_2$ ($\varepsilon_r = 9.6$–13.5), and ZnO ($\varepsilon_r = 7.8$–8.8) (106). Therefore, the high $\mu$ of BLSO crystals may be due to the greatly reduced ionized impurity scattering that results from the enhanced screening strength associated with high $\varepsilon_r$.

### 5.3. Additional Scattering Due to Neutral Impurities in Ba(Sn,Sb)O$_3$

Another important factor for reducing carrier scattering in BLSO is the location of dopants. It is postulated that carrier scattering can be reduced if the dopant is located away from the SnO$_6$ octahedra that compose the main conduction paths. $A$-site dopants would be then favorable for realizing an almost defect-free Sn-O network. To check this postulation, we have systematically studied $n$-dependent $\mu$ behavior in Sb-doped BaSnO$_3$ crystals, as summarized in **Figure 8**. These single crystals were grown by the same flux method as discussed in Section 2.1 by using a mixture of Cu$_2$O and CuO powders (69). In sharp contrast to the case of BLSO, the $\mu$ of BSSO crystals clearly deviates from the $\mu_{ii\_D}$ curve and decreases more steeply than the $\mu_{ii\_D}$ curve with the increase in $n$. Moreover, the measured $n$ in BSSO crystals was only half of the nominal dopant amount, indicating that only half of dopants are activated and the other half of dopants remains as a bound state of Sb$^{5+} + e^-$, which in turn would behave effectively as Sb$^{4+}$ in the scattering process. This result naturally suggests the possible existence of neutral impurity, of which concentration is roughly same as the measured $n$. In conventional semiconductors, the effective mobility due to neutral impurities $\mu_N$ is expressed as

$$\mu_N = \frac{2\pi e}{10 a_B h} \frac{1}{N_N} \text{ with } a_B = \frac{\varepsilon_r \varepsilon_0 h^2}{\pi m^* e^2}, \quad 7.$$

where $N_N$ is the number of neutral impurities per unit volume and $a_B$ is the scaled Bohr radius of a hydrogen-like bound state (111). As the neutral impurities in this case are close to the state of $Sb^{4+}$, we assume that $a_B$ is the ionic radius of the $Sb^{4+}$ ion (0.069 nm), which is taken to be the average radius of $Sb^{3+}$ and $Sb^{5+}$ ions (69). Moreover, we can assume that $N_N$ is roughly same as $n$ to be consistent with the experimental observation. The resultant $\mu_N$ curve based on these assumptions is plotted in **Figure 8**. The $\mu$ values of BSSO crystals are indeed close to the curve of $(\mu_N^{-1} + \mu_{ii\_D}^{-1} + \mu_{ph}^{-1})^{-1}$, indicating that electron scattering is collectively governed by neutral impurities, by ionized dopants, and by phonons. Our results for BSSO crystals thus indicate that carrier scattering becomes more severe in Sn site doping, as the dopants are located in the middle of the $SnO_6$ octahedra, the main conduction path.

### 5.4. Scattering from Dislocations and Other Sources in (Ba,La)SnO₃ Films

In the case of BLSO/SrTiO₃(001) films, additional scattering sources should be considered. $\mu_{ph}$ can again be assumed to be ~600 cm² V⁻¹ s⁻¹ at 300 K. However, in most BLSO/SrTiO₃(001) films, there are numerous threading dislocations (**Figure 4a,b**). Consistent with the presence of a considerable concentration of dislocations $\mu$ of BLSO/SrTiO₃(001) films in **Figure 5** decreased with a decrease in $n$ in the doping range $n \leq 4 \times 10^{20}$ cm⁻³ (32, 61). This phenomenon can be explained by considering threading dislocations as major scattering centers. If the dislocations have an edge component, they introduce acceptor centers along the dislocation line, capturing $n$-type carriers (112, 113). The scattering centers can be smeared out by carrier-induced reduction of a screening length of an isolated charge, resulting in increased $\mu$ with an increase in $n$ ($\mu = \sim n^{1/2}$). $\mu$ of BLSO/SrTiO₃(001) films closely follows the $n^{1/2}$ dependence (**Figure 5**) as observed in conventional semiconductors with threading dislocations (112, 113). When the doping level exceeds $4 \times 10^{20}$ cm⁻³, scattering by dislocations is sufficiently reduced due to enhanced screening by electron carriers, and then scattering by ionized dopants dominates.

In addition, there exist other scattering sources in BLSO films: cation vacancies, cation site mixing, and antisite defects of La impurities. Scanlon (98) argued, on the basis of DFT calculations with a PBE0 hybrid functional, that in undoped BaSnO₃ under an O₂-rich environment, Ba or Sn vacancies can be easily generated. Moreover, the calculated formation energy of La-antisite defects was lower than that of La impurities at Ba sites. The effect of the La-antisite defect in the BLSO system is to trap one electron because La impurities in Sn sites are likely to act as deep acceptors rather than as shallow ones. Those cation vacancies and antisite defects may thus act as additional scattering centers, which may further decrease $\mu$ in BLSO films (81). In sum, $\mu$ in BLSO films can be increased by reducing dislocations and by minimizing cation vacancies or antisite defects of La dopants. With this motivation, there are various research efforts globally to improve $\mu$ in electron-doped BaSnO$_3$ films. Such research includes the use of a buffer layer, substrates with reduced lattice mifit, and state-of-the-art growth techniques like MBE as explained in Section 3.2 and 3.3 above.

# 6. DIRECTIONS FOR VARIOUS DEVICE APPLICATIONS

In this section, we briefly review recent progress in electronic applications and emerging issues based on the BaSnO$_3$ system. Results from our and other research groups have proved the feasibility of many active electronic and optoelectronic devices.

## 6.1. Thin-Film Transistors

Among the diverse forms of oxide electronic devices, perhaps the most fundamental building block of a complex circuitry is a FET, which enables one to control the conductivity of a channel by applying voltage to the gate electrode (125). To demonstrate a thin-film transistor (TFT) employing a BaSnO$_{3-\delta}$ channel, we grew a BaSnO$_{3-\delta}$ thin film by PLD on an insulating BaSnO$_3$(001) substrate to reduce the effects of dislocation-dominant scattering. **Figure 9***e*,*f*,*g* shows the properties of TFTs with the BaSnO$_{3-\delta}$ channel layer. The on-off current ratio increased up to a value of $1.2 \times 10^6$, whereas the field-effect $\mu$ ($\mu_{FE}$) was 48.7 cm$^2$ V$^{-1}$ s$^{-1}$. Other groups have taken the different approach of fabricating a BLSO channel layer on a SrTiO$_3$(001) substrate with a BaSnO$_3$ (**Figure 9***h*,*i*) (79, 115, 126) or (Sr,Ba)SnO$_3$ buffer layer (**Figure 9***j*,*k*) (127). The performance of BaSnO$_3$ TFTs based on the insulating BaSnO$_3$(001) substrate could be improved further with, for example, fine control of defect levels or use of different gate oxides. For detailed explanation of the BaSnO$_{3-\delta}$ TFTs in **Figure 9**, which were constructed on a BaSnO$_3$(001) substrate, and of the TFT performance of other research groups using the buffer layer/SrTiO$_3$(001) substrate, please refer to the **Supplemental Material**.

## 6.2. *pn* Junctions

With the superior thermal stability of oxygen in BaSnO$_3$, as described in Section 4.1, *p*-type BaSnO$_3$, once realized, should have less degradation than, e.g., the degradation due to oxygen vacancies at the interface. To create *p*-type TCOs and TOSs in the BaSnO$_3$ system, both Ba sites and Sn sites can be doped by acceptor ions. Our group searched possible acceptors in BaSnO$_3$ and found that Co$^{3+}$ at Sn sites may be a viable *p*-type dopant in BaSnO$_3$. We found that BaSn$_{0.9}$Co$_{0.1}$O$_3$ films grown on BaSn$_{0.97}$Sb$_{0.03}$O$_3$ single crystals, *p*-BaSn$_{0.9}$Co$_{0.1}$O$_3$/*n*-BaSn$_{0.97}$Sb$_{0.03}$O$_3$ diodes, exhibit a rectification ratio of 2.9 × 10$^3$ (±2 V) at room temperature (**Figure 10*a,b***). We found a similar rectification ratio even after thermal annealing at 200°C. This result and recent reports of *pn* diodes made of *p*-(Ba,K)SnO$_3$ films (**Figure 10*e,f***)(128) indicate that thermally stable *pn* junctions can be fabricated from *p*-type TOSs based on the BaSnO$_3$ system (by either Ba or Sn site doping), increasing the application potential of perovskite stannates. For more information of *pn*-diodes on **Figure 10**, please refer to the **Supplemental Material**.

### 6.3. Solar Cells and Photoconductors

Other promising applications are solar cells and photoconductors. Although these areas have great potential, there are only limited reports (129-133). We summarize recent research in the **Supplemental Material**.

### 7. POTENTIAL TWO-DIMENSIONAL ELECTRON GASES IN BARIUM STANNATE-BASED HETEROSTRUCTURES

From the viewpoints of both fundamental physics and new device applications, one of the most exciting developments in oxide semiconductors is the realization of 2DEGs at heterointerfaces. Such oxide interfaces can be a platform for observing a plethora of emergent and novel electronic states that have not been observed in any of the constituent materials. An archetypal example is the 2DEGs realized at the LaAlO$_3$-SrTiO$_3$ heterointerface (134). However, the underlying mechanism remains unclear (135) because SrTiO$_3$ can easily deviate from the oxygen stoichiometry (86, 136, 137). The study of 2DEGs at the heterointerface of a BaSnO$_3$ layer with a polar oxide could be useful, as BaSnO$_3$ has both the thermal stability of oxygen and a low diffusion coefficient of oxygen (66). Moreover, BaSnO$_3$ has the same cubic perovskite structure as SrTiO$_3$, which could allow greater opportunities to grow heteroepitaxial films with other perovskite materials. Therefore, BaSnO$_3$ may be another exciting way to realize 2DEGs at heterointerfaces formed with other insulators.

### 7.1. Theoretical Predictions for Two-Dimensional Electron Gases at Heterointerfaces

There have been theoretical discussions on possible 2DEGs at the heterojunction of BaSnO$_3$ with (*a*) KTaO$_3$ or LaAlO$_3$ (138) and (*b*) KTaO$_3$ or KNbO$_3$ (139). In 2016, Krishnaswamy et al. (97) detailed two major strategies for the realization of 2DEGs in the BaSnO$_3$ system: polar discontinuity and modulation doping. One key requirement in both strategies is to find an oxide material with a higher conduction band minimum than BaSnO$_3$ such that the electron carriers are well confined to the BaSnO$_3$ layer. Under this condition, free electrons will readily spill over into, and remain in, the BaSnO$_3$ layer. As the two strategies are important for future research, we here further explain the calculation results by Krishnaswamy et al. and the related physics.

Between these two main strategies, doping due to polar discontinuity is particularly appealing, as it is a form of remote doping without the direct introduction of dopants or chemical disorder. For polar discontinuity doping, like the polar mechanism invoked for the LaAlO$_3$/SrTiO$_3$(001) system (134), a thin film of an $A^{3+}B^{3+}(O^{2-})_3$ polar oxide with relatively good lattice match, such as LaInO$_3$, on the SnO$_2$-terminated (001) surface of a BaSnO$_3$ single crystal (nonpolar oxide) could lead to the formation of 2DEGs (**Figure 11a**). As long as the band offset of the polar material is high enough not to spill over the carriers at the interface, the ideal 2D carrier density $n_{2D,ideal}$ expected at the interface should be 0.5 electrons per in-plane unit cell, which corresponds to $n_{2D,ideal} = 2.9 \times 10^{14}$ cm$^{-2}$.

In modulation doping (**Figure 11b**), dopants are introduced in a barrier material whose conduction band minimum is higher than that of BaSnO$_3$. Electrons from the dopants can transfer from the barrier material to the BaSnO$_3$ layer to form 2DEGs at the interface. An insulating spacer layer should be grown to enhance the separation between electrons and ionized donors. Those 2DEGs are then separated from the dopants and is therefore less prone to impurity scattering (140, 141). The most viable dopant profile may be delta doping, whereby the dopants are introduced as a sheet in the barrier material at a distance *d* away from the interface (142, 143). Krishnaswamy et al. (97) calculated $n_{2D}$ as a function of dopant sheet concentrations on the basis of the band offsets of the barrier materials (SrTiO$_3$, LaInO$_3$, and KTaO$_3$). The maximum $n_{2D}$ between BaSnO$_3$ and other $ABO_3$ $n_{2D,max}$ induced in the BaSnO$_3$ layer generally increases as the conduction band offsets (CBOs) between BaSnO$_3$ and barrier materials increase.

**Figure 11***c* replots the calculation results of Krishnaswamy et al. (97), summarizing the theoretically used CBOs and the resultant $n_{2D,max}$ between BaSnO$_3$ and other $ABO_3$. Moreover, $a_c$ and $a_{pc}$ of the barrier materials, and their lattice misfits with BaSnO$_3$ are also plotted. Here, $n_{2D,max}$ was extracted from an $n_{2D}$-versus-CBO plot when 0.5 electrons per in-plane unit cell were introduced (figure 5 in Reference 97). The CBOs used in these calculations were generally larger than those from the experimental results (**Figure 12**) because the calculations employed a smaller band gap = 2.4 eV of BaSnO$_3$ from a HSE06 hybrid functional, whereas the experimental $E_g$ is ~3.0 eV. In turn, the predicted $n_{2D,max}$ is expected to be higher than that from the real experimental data.

### 7.2. Experimental Situation

Whereas there are ample predictions from theory, the related experimental efforts on 2DEGs in BaSnO$_3$ are still nascent. A necessary step is to grow a heteroepitaxial film in which BaSnO$_3$ works as a channel material and the other joint material as a barrier material. For this purpose, the candidate barrier material should have (*a*) a higher conduction band minimum than does BaSnO$_3$ and (*b*) a good lattice match with BaSnO$_3$. Experiments should be undertaken to determine the CBOs between the BaSnO$_3$ and the candidate materials. The use of accurate $E_g$ or band offsets can provide more accurate estimation of $n_{2D}$ at the interface.

Chambers et al. (144) recently reported, using photoemission spectroscopy, that the conduction band minimums of SrTiO$_3$ and LaAlO$_3$ are higher than the conduction band minimum of BaSnO$_3$ by 0.42 and 3.72 eV, respectively. These authors thus discussed the possibility of 2DEGs at the SrTiO$_3$-BaSnO$_3$ interface by modulation doping or at the LaAlO$_3$-BaSnO$_3$ interface by polar discontinuity doping. In addition to this research, it is worthwhile to list promising candidate materials on the basis of their band alignments. **Figure 12** presents the summary of experimental information on the band alignments of candidate materials to provide some guidance for future research. For details on how **Figure 12** was generated, please refer to the **Supplemental Material**.

In 2016, Kim et al. (116) reported a significant increase in conductivity after LaInO$_3$ films were grown on top of BLSO films (0.1% < La < 0.5%), which in turn were grown on a BaSnO$_3$ buffer (110 nm)/SrTiO$_3$(001) substrate. In contrast, there was no increase in the case of undoped films. The authors claimed that the conductivity increase is due to the formation of a conducting interface by polar discontinuity doping at the LaInO$_3$-BLSO interface and that such an increase can occur only when the $E_F$ of BaSnO$_3$ increases due to (low) La doping. However, the dislocations in the low-La-doped BaSnO$_3$ film may have decreased the original $E_F$ to make the film $p$ type. Therefore, whether the 2DEGs actually formed at the polar interface between insulating BaSnO$_3$ and LaInO$_3$ awaits more extensive exploration.

To summarize the experimental situation, the growth of large-area and homogeneous thin films without dislocations or defects by the use of, e.g., MBE with a proper substrate like BaSnO$_3$ will be necessary for elucidating the interfacial physics in the BaSnO$_3$-based heterostructure. For this, it will be useful to realize a BaSnO$_3$(001) substrate with an atomically flat surface and a terrace-step structure. We have succeeded in obtaining such a structure at the surface of a highly insulating BaSnO$_3$(001) substrate, and experimental details are expected to be published in the future (W.-J. Lee & K. H. Kim, unpublished manuscript). **Figure 11***d* shows such a terrace-step structure with a step height of 4.1 Å at the surface, as obtained from a surface etching and thermal annealing procedure. This development will allow researchers to realize 2DEGs at the interfaces of BaSnO$_3$ layers in perovskite stannates.

## 8. CONCLUDING PERSPECTIVES AND OUTLOOK

Above we review recent progress on the newly emerging electron-doped BaSnO$_3$ system as an excellent example of a TOS and TCO, explaining this system's physical properties, device application potential, and scientific challenges. The BaSnO$_3$ system has two unique physical properties that makes it superior to other oxides. The first is the high $\mu =$ ~320 cm$^2$ V$^{-1}$ s$^{-1}$ at room temperature, which is essential for the fast operation of electronic devices. The second is excellent thermal stability, by which one can potentially overcome, e.g., interface degradation and/or persistent current problems in oxide *pn* junctions and transistors. In addition, its cubic perovskite structure is versatile enough to form useful heterointerfaces with various other perovskite oxides and allows for ionic dopant substitutions into both Ba and Sn sites and thus for flexible *p* and *n* doping. This structure also allows for lattice engineering to tune $E_g$ and band alignment. Therefore, with regard to the above physical properties, the BaSnO$_3$ system bears greater similarity to traditional compound semiconductors than to the usual oxides.

We also discuss above that, on the basis of those superior properties, TFTs with high field-effect mobility and *pn* junctions with a high rectification ratio have been developed. Due to such progress as well as the thermal stability of BaSnO$_3$ with respect to oxygen, we expect the BaSnO$_3$ system to provide an unprecedented opportunity to realize high-temperature, high-power, functional semiconductor devices. Before such devices can be commercialized, however, numerous scientific issues should be resolved. In view of the research history of conventional semiconductors, the ability to control dopant concentration in the nondegenerate regime seems to be necessary. The combination of proper substrates and growth methods such as MBE should reduce dislocations and impurities to allow for the control of dopant concentration. The high mobility of the BaSnO$_3$ system is likely to be further enhanced in engineered oxide heterostructures grown by MBE.

Various potential applications and scientific issues confront researchers. **Figure 13** plots the expected $\mu$ behavior of the electron-doped BaSnO$_3$ system with available experimental data at 2 K as a function of *n*. The expected $\mu$ was calculated from an ionized impurity scattering model in a degenerate doping regime. *AB*O$_3$-BaSnO$_3$ interfaces can hold 2DEGs by either modulation or polar discontinuity doping, thereby paving the way for the investigation of quantum phenomena. First, the number of phonons at such a low temperature will be reduced, resulting in almost zero phonon scattering. Then, one can investigate $\mu$ in a low-*n* region ($n < 10^{18}$ cm$^{-3}$). According to the ionized impurity scattering model, $\mu$ increases with a decrease in *n* due to the decreased number of scattering centers. In this regime, one can observe the quantum Hall effects in the 2DEGs at *AB*O$_3$-BaSnO$_3$ heterostructures, which allowed for understanding of the fundamental properties of the 2DEGs at LaAlO$_3$-SrTiO$_3$ heterostructures (141, 143, 145). Therefore, by employing modulation or polar discontinuity doping, quantum Hall effects in the BaSnO$_3$ system may well be discovered.

Another fascinating phenomenon, superconductivity in the $BaSnO_3$ system, is likely to occur in a high-$n$ regime. On the basis of carrier accumulation by electrical double-layer (EDL) transistors as well as modulation and polar discontinuity doping, interfacial superconductivity recently emerged in channels made of similar wide-band-gap oxides, including $SrTiO_3$, $KTaO_3$, and the $LaAlO_3$-$SrTiO_3$ heterointerface (142, 146-148). Therefore, we also anticipate that $BaSnO_3$-based heterostructures will exhibit superconductivity by the accumulation of carriers by EDL transistors and/or modulation or polar discontinuity doping.

## DISCLOSURE STATEMENT

The authors are not aware of any affiliations, memberships, funding, or financial holdings that might be perceived as affecting the objectivity of this review.

## ACKNOWLEDGMENTS

This work was financially supported by National Creative Research Initiative (2010-0018300); by the Korea-Taiwan Cooperation Program (0409-20150111); and by Global Collaborative Research Projects (2016K1A4A3914691) through NRF of Korea. We appreciate experimental help from Jin Hyeok Kim and stimulating discussions with Takhee Lee, Seunghun Hong, and Philip Kim.

## LITERATURES

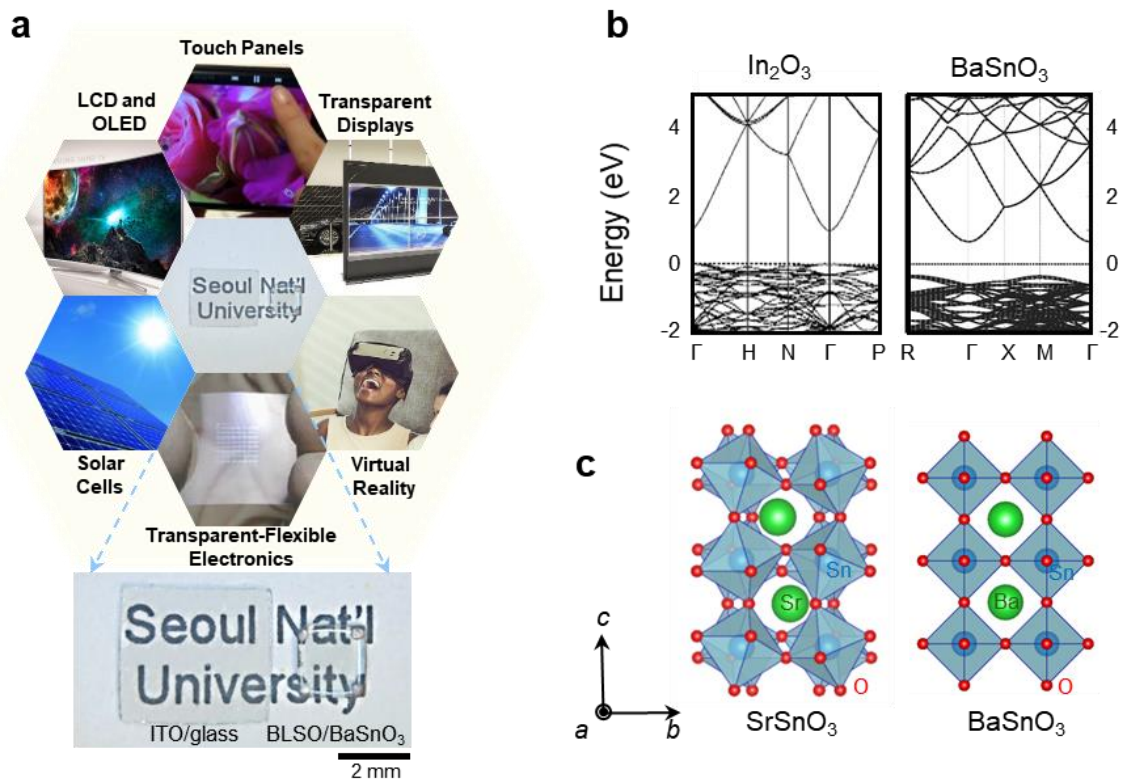

Figure 1 (*a*) Varieties of industrial applications employing transparent conducting oxides and transparent oxide semiconductors. The central photograph and its expanded view represent a transparent $(In,Sn)_2O_3$ (ITO) film on a glass substrate and a $(Ba,La)SnO_3$ (BLSO) film on a $BaSnO_3(001)$ substrate. (*b*) Band structures of $In_2O_3$ and $BaSnO_3$. (*c*) Crystal structures of $BaSnO_3$ and $SrSnO_3$ with cubic and orthorhombic perovskites, respectively. Other abbreviations: LCD, liquid-crystal display; OLED, organic light-emitting diode. An Image of transparent-flexible electronics in panel *a* adapted with permission from Reference 6; copyright 2004, Nature Publishing Group. Images of LCD and OLED, touch panels, transparent displays, and virtual reality in panel a from Reference 12; copyright 2016, Samsung Electronics. An image of solar cells in panel *a* adapted with permission from http://www.slashgear.com/; copyright 2016, SlashGear. A band structure of $In_2O_3$ in Panel *b* adapted with permission from Reference 35; copyright 2001, American Physical Society. A band structure of $BaSnO_3$ in panel *b* adapted with permission from Reference 61; copyright 2012, American Physical Society. Panel *c* was drawn on our own for this manuscript by Diamond software.

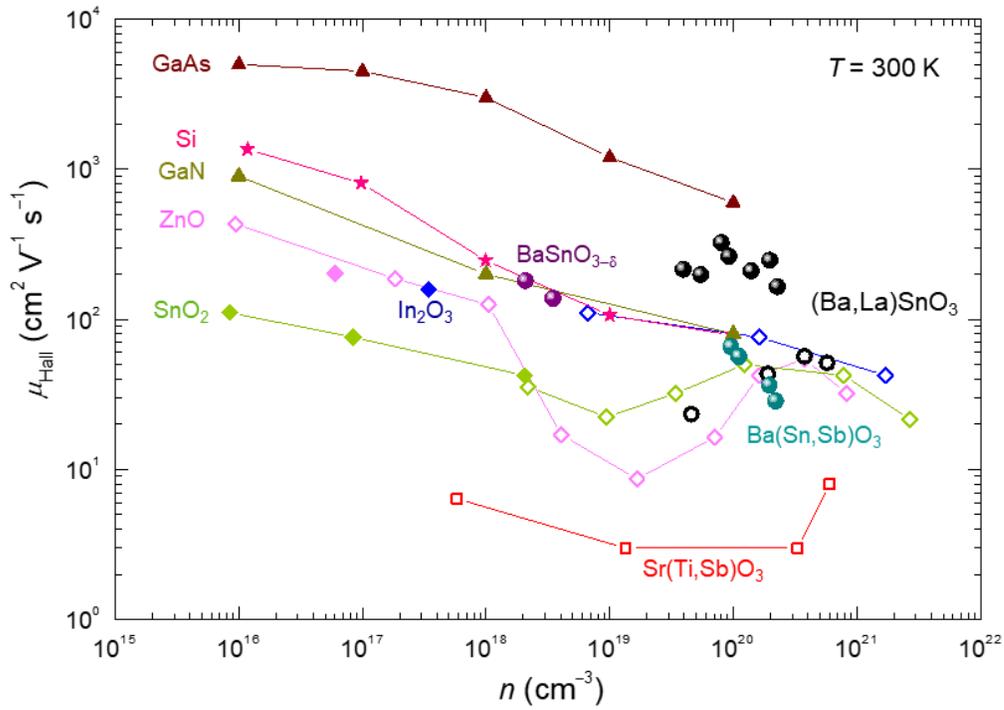

Figure 2 A summary of electron mobility $\mu$ versus electron carrier concentration $n$ or electron-doped $BaSnO_3$ systems, typical transparent conducting oxides and transparent oxide semiconductors, and conventional semiconductors at 300 K. Solid and open symbols represent data for single crystals and thin films, respectively. Original data of $(Ba,La)SnO_3$ ([61]), $Ba(Sn,Sb)O_3$ ([69]), $BaSnO_{3-\delta}$ ([69]), $In_2O_3$ ([70]), $SnO_2$ ([70]), ZnO ([21]), $Sr(Ti,Sb)O_3$ ([71]), Si ([72]), GaN ([73]), and GaAs ([74]) were replotted.

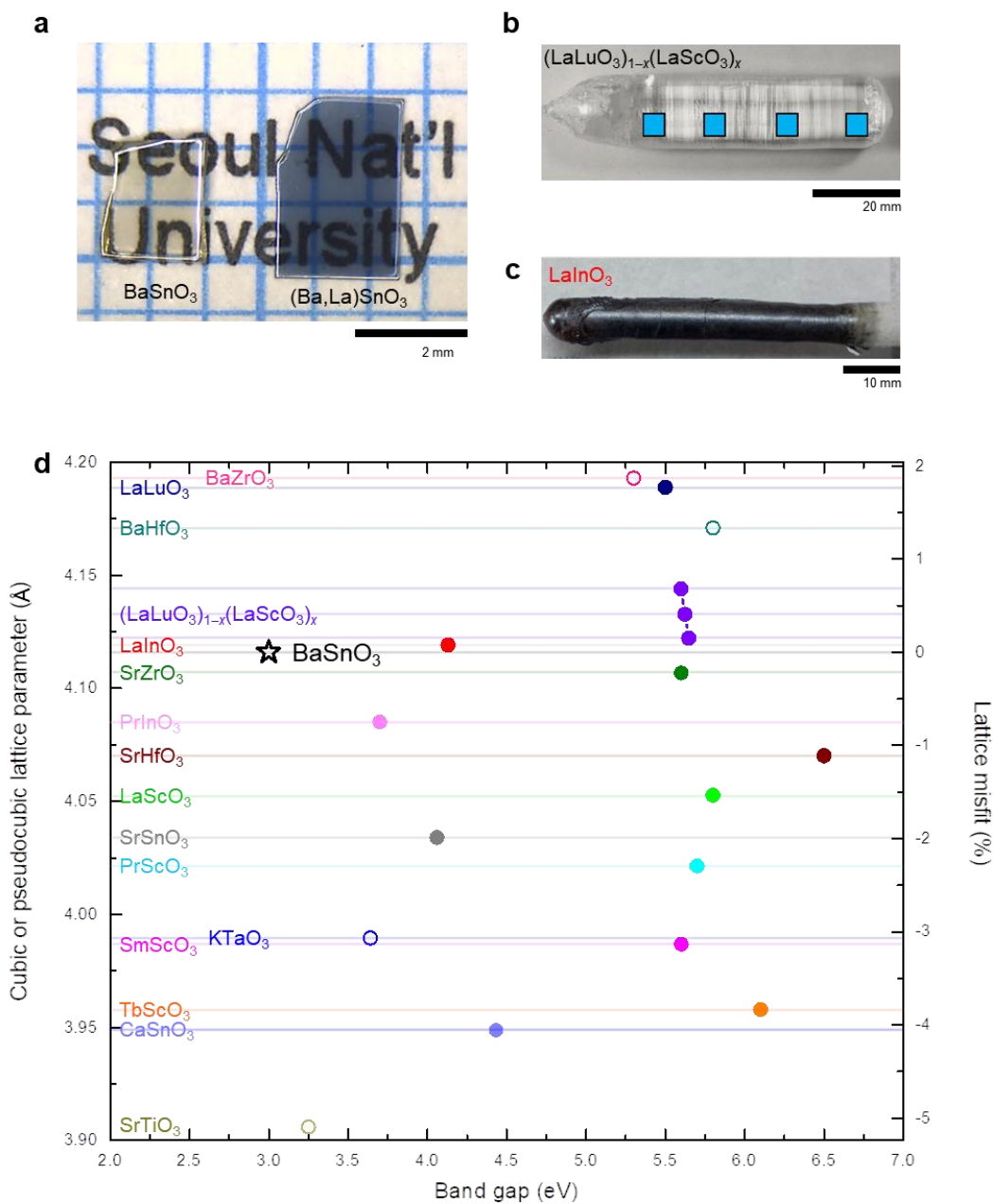

Figure 3 (*a*) Transparent BaSnO$_3$ (700 μm thick) and (Ba,La)SnO$_3$ (60 μm thick) single crystals grown by the flux method. (b) A (LaLuO$_3$)$_{1-x}$(LaScO$_3$)$_x$ single crystal grown by the Czochralski method (114). (*c*) A LaInO$_3$ single crystal grown by the floating zone technique (149). (*d*) Band gap versus either $a_c$ or $a_{pc}$ (and lattice misfit with BaSnO$_3$) in cubic (*open*) or orthorhombic (*solid*) perovskites. Most of the band gap values are from experiments on optical gaps, whereas those of scandates are from first-principles calculations. Supplemental Table 1 includes detailed data and references. Panel *b* adapted with permission from Reference (114); copyright 2016, Elsevier.

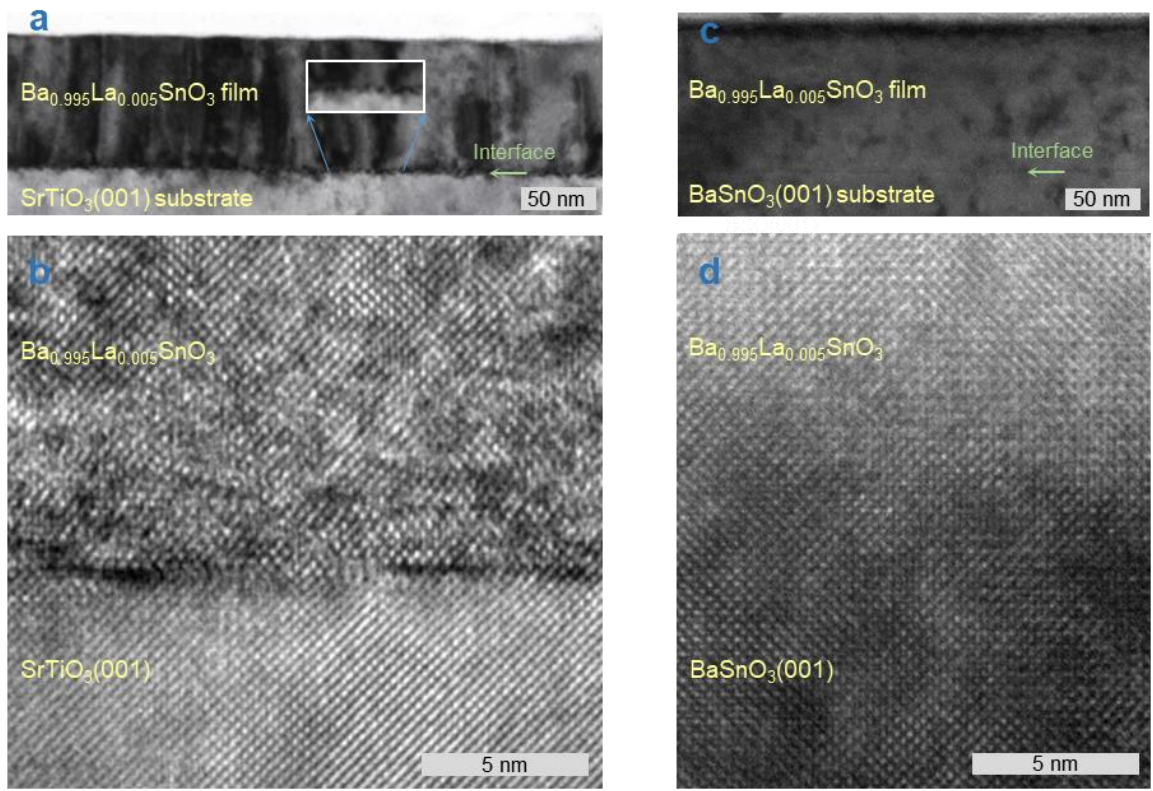

Figure 4 TEM images showing film-substrate interfaces. Panels *a* and *b* are low- and high-magnification cross-sectional TEM images, respectively, of the $Ba_{0.995}La_{0.005}SnO_3$-$SrTiO_3$(001) interface. Panel *c* and *d* are images of the $Ba_{0.995}La_{0.005}SnO_3$-$BaSnO_3$(001) interface. Panels *c* and *d* adapted with permission from Reference (81); copyright 2016, AIP.

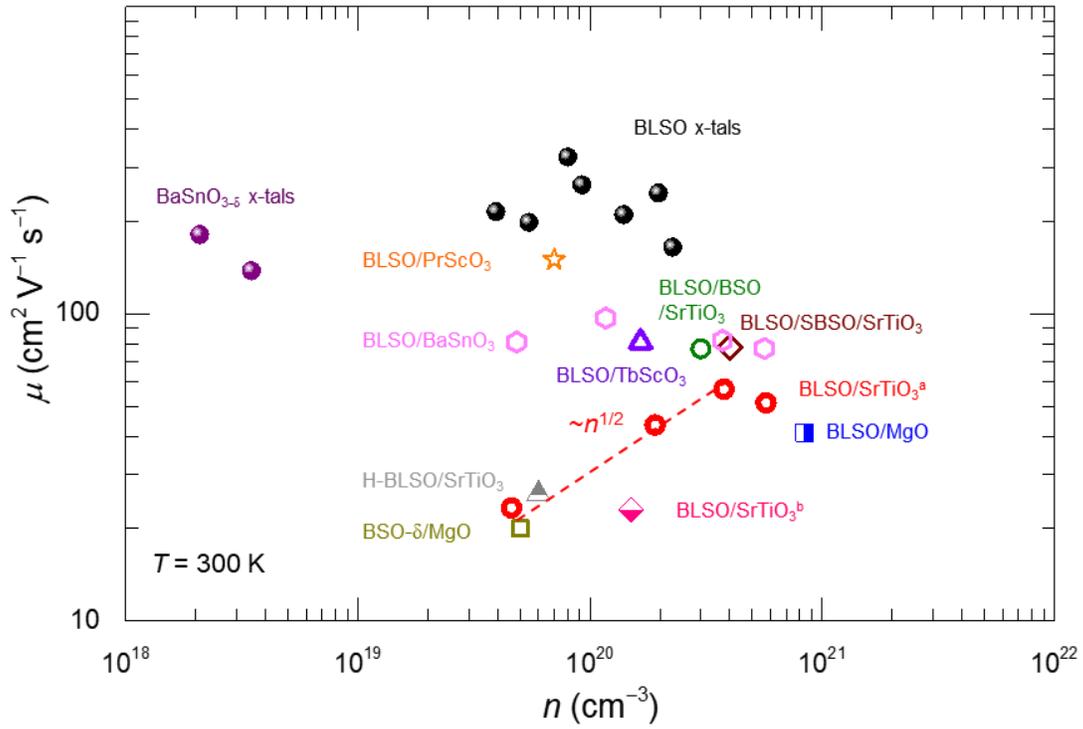

Figure 5 Summary of experimental $\mu$ data in thin films prepared by pulsed laser deposition [BLSO/SrTiO$_3$$^a$ (61, 119), BLSO/SmScO$_3$ (77), BLSO/BSO/SrTiO$_3$ (79), BLSO/SBSO/SrTiO$_3$ (80), BLSO/BaSnO$_3$ (81), H-BLSO/SrTiO$_3$ (101), (Ba,Gd)SnO$_3$/SrTiO$_3$ (118), BLSO/MgO (120)], sputtering [BaSnO$_{3-\delta}$/MgO (122)], solution deposition [BLSO/SrTiO$_3$$^b$ (123)], and molecular beam epitaxy [BLSO/PrScO$_3$ (31), BLSO/TbScO$_3$ (78)] at room temperature. $\mu$ of BLSO/SrTiO$_3$$^a$ (61) closely follows the $n^{1/2}$ dependence (red dashed line). Abbreviations: $\mu$, electron mobility; $n$, electron carrier concentration; BSO, BaSnO$_3$; BLSO, (Ba,La)SnO$_3$; H-BLSO, hydrogen-treated BLSO; SBSO, (Sr,Ba)SnO$_3$.

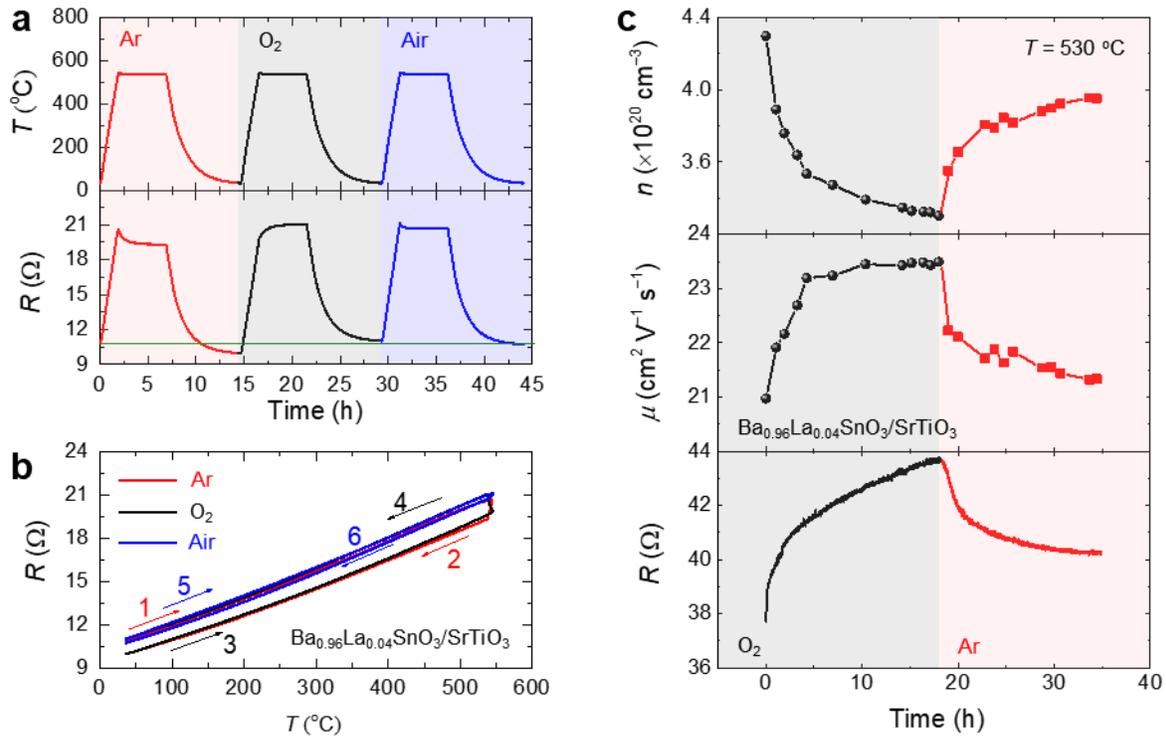

Figure 6 Thermal stability of the BaSnO$_3$ system. (*a*) Temperature *T* and gas atmosphere were varied according to the profile in the top panel. The resultant resistance *R* variation of a Ba$_{0.96}$La$_{0.04}$SnO$_3$/SrTiO$_3$(001) film is plotted in the bottom panel. (*b*) *R* decreased (increased) by approximately 8% (9.5%) under Ar (O$_2$) atmosphere in 5 h at 530°C, whereas it changed by only approximately 1.7% in air. The numbers (1 through 6) along with their corresponding arrows denote the sequence of measurements under Ar (1, 2), O$_2$ (3, 4), and Air (5, 6) atmospheres. (*c*) Electron carrier concentrations *n*, electron mobility *μ*, and *R* as extracted from time-dependent Hall effect measurements in the Ba$_{0.96}$La$_{0.04}$SnO$_3$/SrTiO$_3$(001) film at 530°C under O$_2$ or Ar atmosphere. Panels *a* and *b* adapted with permission from Reference 32; copyright 2012, The Japan Society of Applied Physics. Panel *c* adapted with permission from Reference 66; copyright 2015, John Wiley and Sons.

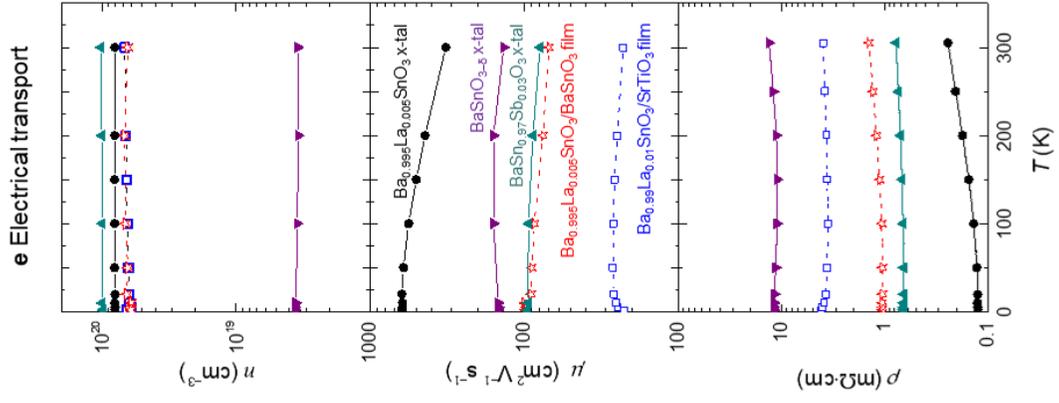

**a Electronic structure**

$Ba_{0.963}La_{0.037}SnO_3$

$BaSn_{0.963}Sb_{0.037}O_3$

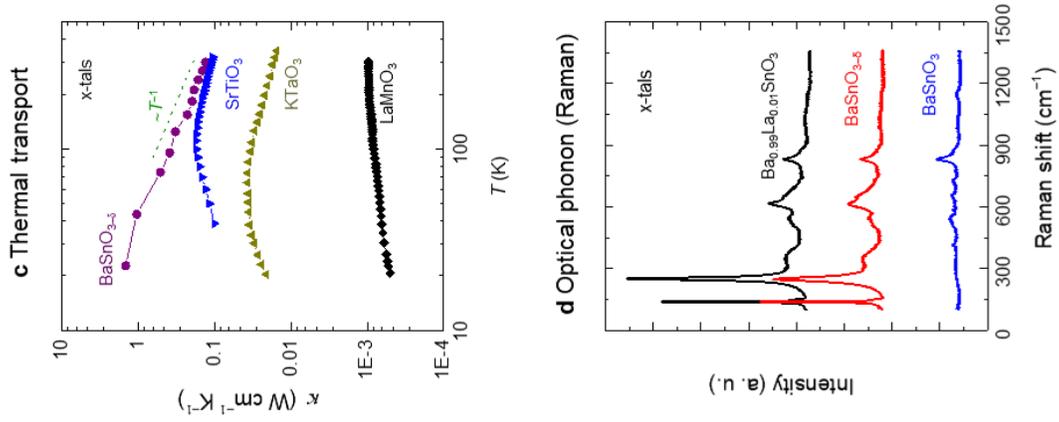

**b Optical properties**

**c Thermal transport**

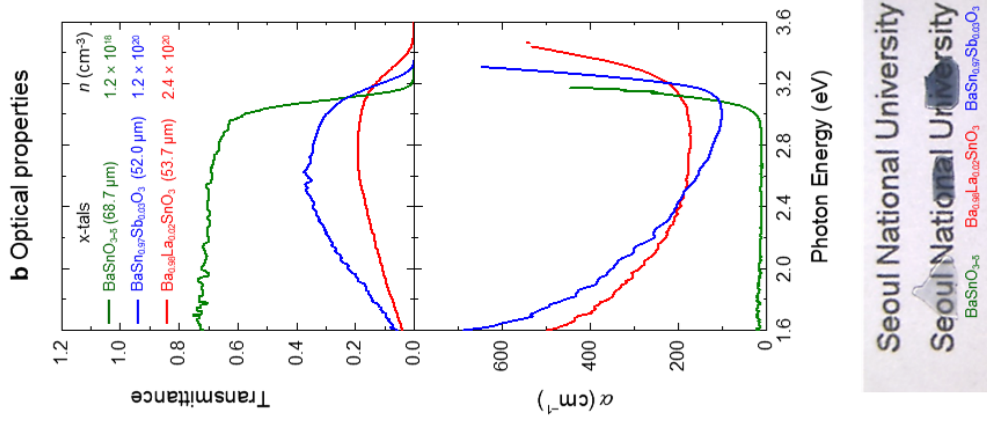

**d Optical phonon (Raman)**

**e Electrical transport**

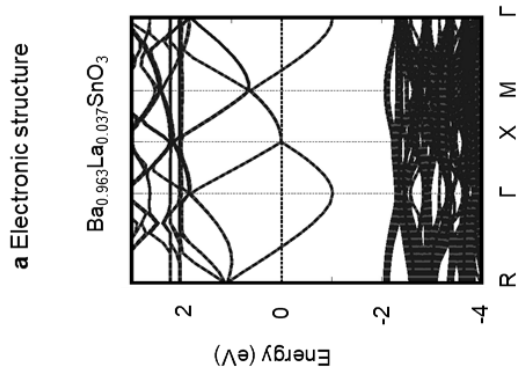

Figure 7 Various physical properties of the electron-doped $BaSnO_3$ system. (*a*) Band structures of $Ba_{0.963}La_{0.037}SnO_3$ and $BaSn_{0.963}Sb_{0.037}O_3$ from first-principles calculations with $3 \times 3 \times 3$ unit cells. (*b*) Transmittance and absorption coefficient $\alpha$ of $BaSnO_{3-\delta}$, $BaSn_{0.97}Sb_{0.03}O_3$, and $Ba_{0.98}La_{0.02}SnO_3$ single crystals (x-tals). The photograph (*bottom*) shows polished $BaSnO_{3-\delta}$, $BaSn_{0.97}Sb_{0.03}O_3$, and $Ba_{0.98}La_{0.02}SnO_3$ single crystals. (*c*) Thermal conductivity $\kappa$ of $BaSnO_{3-\delta}$ (89), $SrTiO_3$ (90), $KTaO_3$ (91), and $LaMnO_3$ (92) single crystals. (*d*) Raman spectra of $Ba_{0.99}La_{0.01}SnO_3$, $BaSnO_{3-\delta}$, and $BaSnO_3$ single crystals, providing evidence that the presence of electron carriers produces local lattice distortions, thereby locally breaking the cubic symmetry. (*e*) Electrical transport properties: Electron carrier concentration $n$, electron mobility $\mu$, and electrical resistivity $\rho$ for selected $Ba_{0.995}La_{0.005}SnO_3$ (61), $BaSn_{0.097}Sb_{0.03}O_3$ (69), and $BaSnO_{3-\delta}$ single crystals (69) and $Ba_{0.995}La_{0.005}SnO_3/BaSnO_3(001)$ and $Ba_{0.99}La_{0.01}SnO_3/SrTiO_3(001)$ (61) thin films. A band structure of $Ba_{0.963}La_{0.037}SnO_3$ in panel *a* adapted with permission from Reference (61); copyright 2012, American Physical Society. A band structure of $BaSn_{0.963}Sb_{0.037}O_3$ in panel *a* adapted with permission from Reference (69); copyright 2013, American Physical Society. Optical properties of $BaSnO_{3-\delta}$ and $Ba_{0.98}La_{0.02}SnO_3$ in panel *b* adapted with permission from Reference (61); copyright 2012, American Physical Society. Optical properties of $BaSn_{0.97}Sb_{0.03}O_3$ in panel *b* adapted with permission from Reference (69); copyright 2013, American Physical Society. $\kappa$ of $BaSnO_{3-\delta}$ in panel *c* adapted with permission from Reference (89); copyright 2014, Elsevier. $\kappa$ of $SrTiO_3$ in panel *c* adapted with permission from Reference (90); copyright 2008, AIP. $\kappa$ of $KTaO_3$ in panel *c* adapted with permission from Reference (91); copyright 2008, AIP. $\kappa$ of $LaMnO_3$ in panel *c* adapted with permission from Reference (92); copyright 1997, American Physical Society. Electrical transport of $Ba_{0.995}La_{0.005}SnO_3$ x-tal and $Ba_{0.99}La_{0.01}SnO_3/SrTiO_3$ film in panel *e* adapted with permission from Reference (61); copyright 2012, American Physical Society. Electrical transport of $BaSnO_{3-\delta}$ and $BaSn_{0.97}Sb_{0.03}O_3$ x-tal in panel *e* adapted with permission from Reference (69); copyright 2013, American Physical Society.

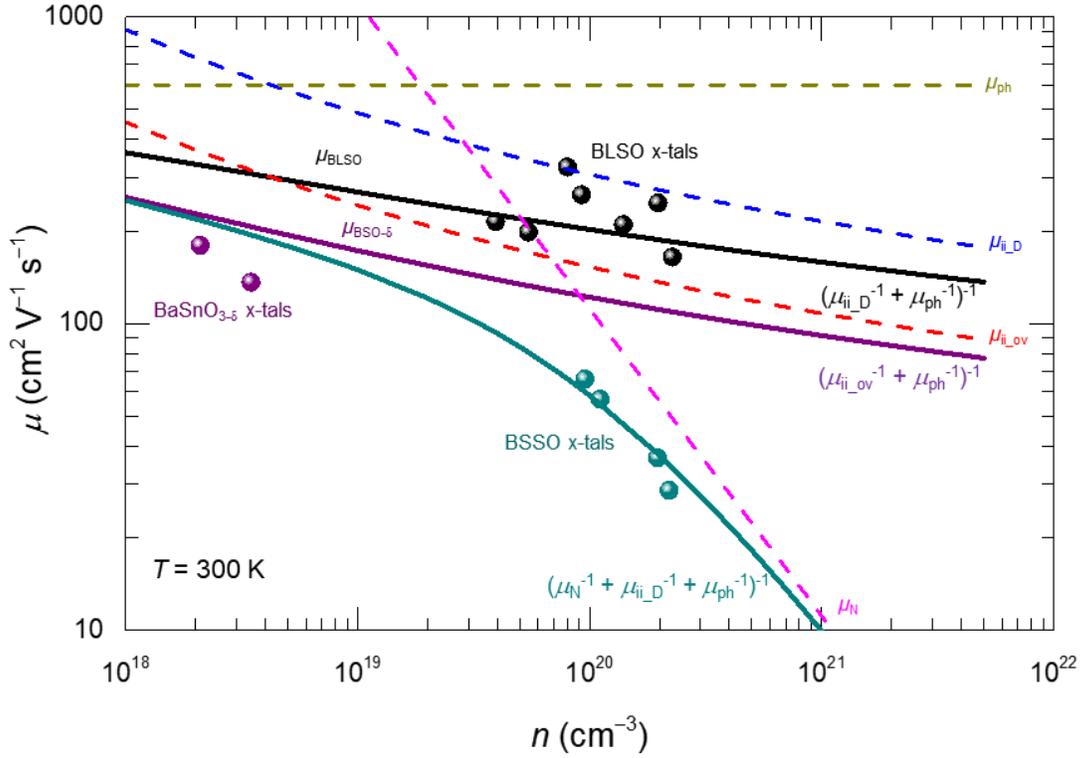

Figure 8 The experimental $\mu$ data of (Ba,La)SnO$_3$ (BLSO), Ba(Sn,Sb)O$_3$ (BSSO), and BaSnO$_{3-\delta}$ single crystals (x-tals) are plotted as solid circles (61, 69). Theoretically estimated $\mu$ curves from individual scattering sources of phonons ($\mu_{ph}$), ionized La or Sb dopants ($\mu_{ii\_D}$), oxygen vacancies ($\mu_{ii\_ov}$), and neutral impurities ($\mu_N$) are plotted as dashed lines. The solid lines refer to the estimated $\mu$ from the two or three scattering sources. Abbreviations: $\mu$, electron mobility; $n$, electron carrier concentration.

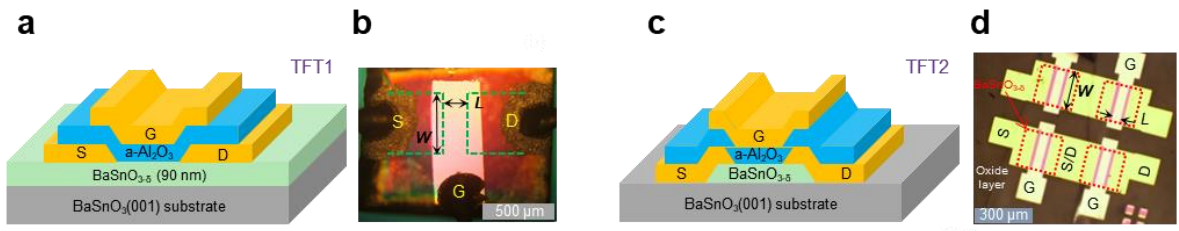

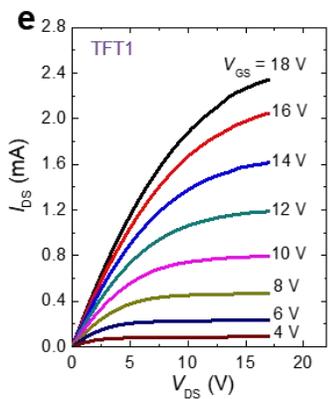 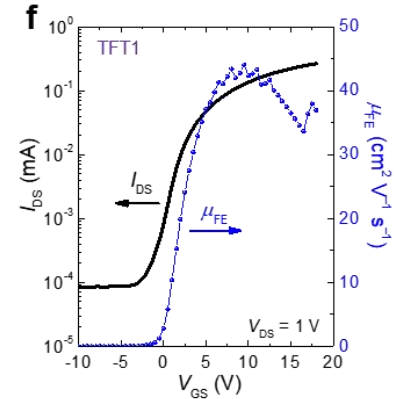 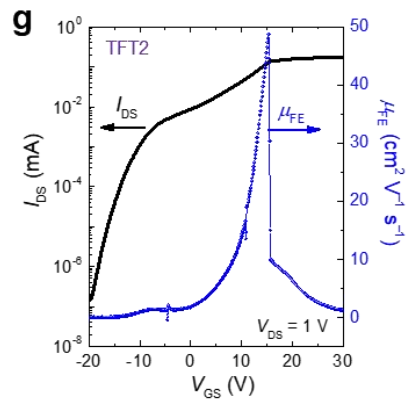

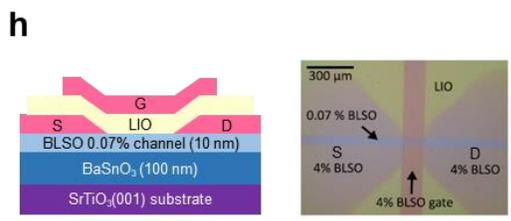 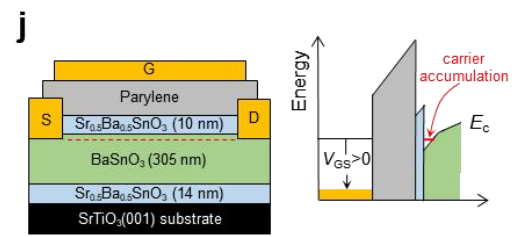

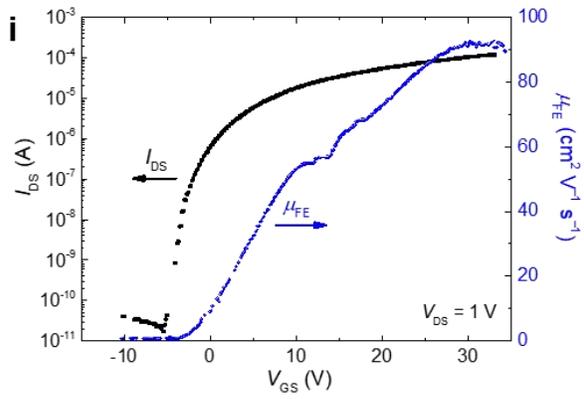 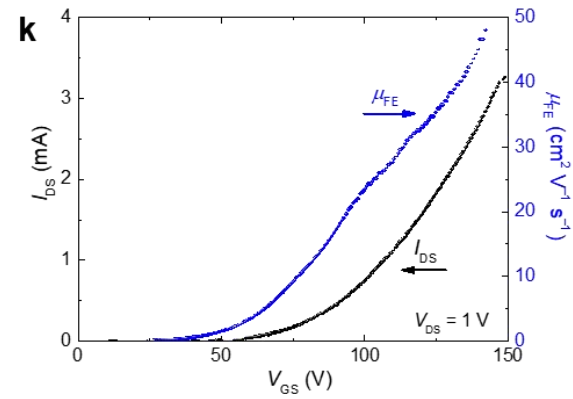

Figure 9 Thin-film field-effect transistors (TFTs) with a BaSnO$_{3-\delta}$ channel layer. (*a*) A schematic device structure for BaSnO$_{3-\delta}$ TFT/BaSnO$_3$(001) (TFT1). (*b*) A top view of TFT1 under an optical microscope. (*c*) A schematic device structure and (*d*) a top view of an image showing the four devices of another type of BaSnO$_{3-\delta}$ TFT (TFT2) fabricated by use of Si stencil masks. The channel length (*L*) and width (*W*) of TFT2 are 50 and 200 μm, respectively. (*e*) Output ($I_{DS}$ versus $V_{DS}$) and (*f*) transfer ($I_{DS}$ versus $V_{GS}$) characteristics of TFT1 and (*g*) transfer characteristics of TFT2. (*h*) Schematic device structure of BLSO-TFT with LIO gate dielectric (*left panel*). The top view of BLSO-TFT by an optical microscope (*right panel*). (*i*) Transfer characteristics of BLSO-TFT. (*j*) Schematic of the structure of SBSO/BSO-TFT (*left panel*). Schematic of conduction band diagram for SBSO/BSO-TFT (right panel). Accumulated carriers at SBSO/BSO. (*k*) Transfer characteristic of SBSO/BSO-TFT. Other abbreviations: D, drain; FE, field effect; G, gate; S, source; BSO, BaSnO$_3$; LIO, LaInO$_3$; SBSO, Sr$_{0.5}$Ba$_{0.5}$SnO$_3$. Panels *h* and *i* adapted with permission from Reference (115); copyright 2016, AIP. Panels *j* and *k* adapted with permission from Reference (127); copyright 2016, AIP.

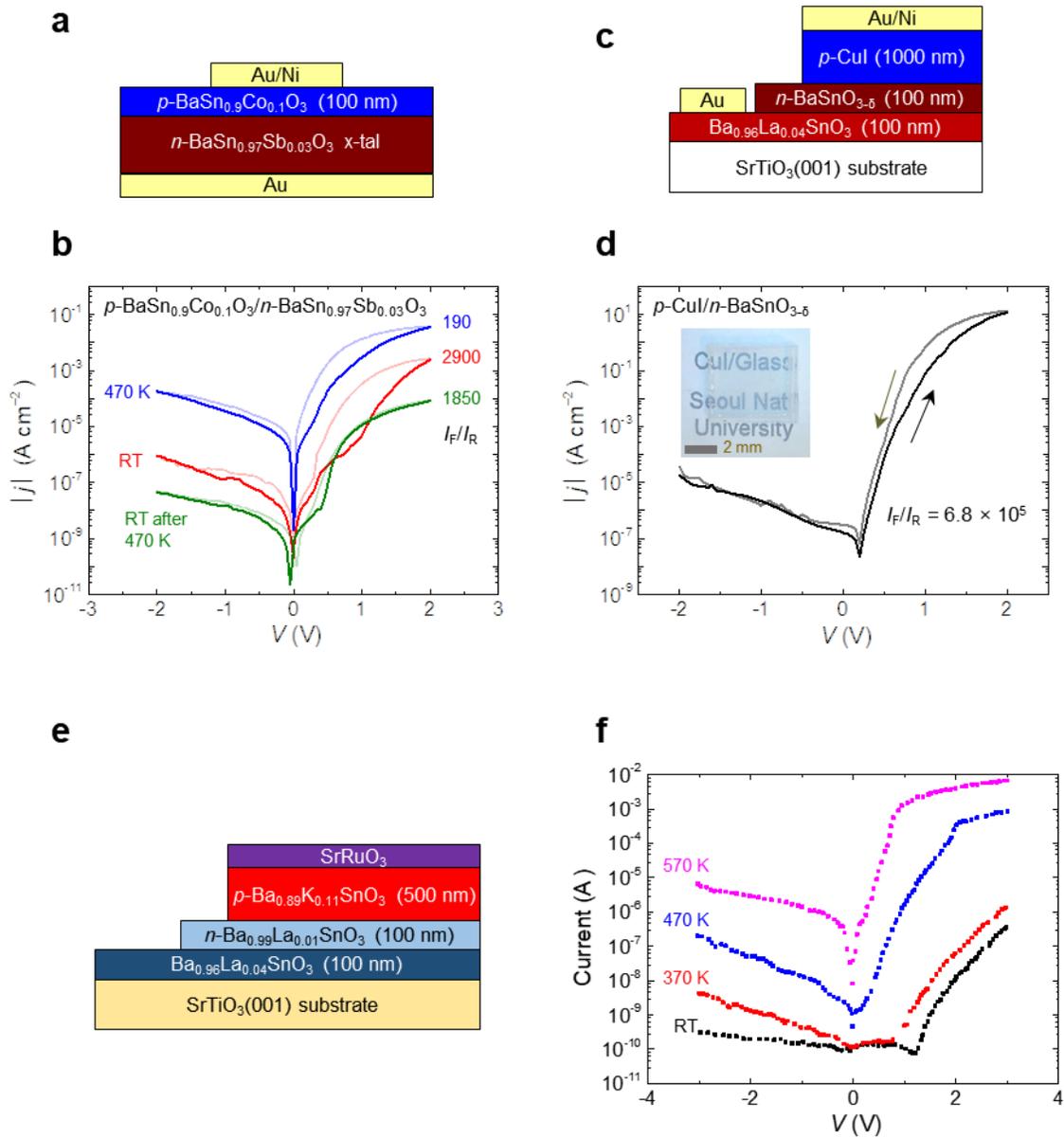

Figure 10 *pn* diodes based on the doped BaSnO$_3$ system. (*a,c*) Schematic device structures of diodes having junctions made of (*a*) BaSn$_{0.9}$Co$_{0.1}$O$_3$ films/BaSn$_{0.97}$Sb$_{0.03}$O$_3$ x-tal and (*c*) CuI/BaSnO$_{3-\delta}$ thin films on SrTiO$_3$(001) substrates (152). (*b,d*) Current density–versus–voltage curves of (*b*) BaSn$_{0.9}$Co$_{0.1}$O$_3$/BaSn$_{0.97}$Sb$_{0.03}$O$_3$ diodes and (*d*) CuI/BaSnO$_{3-\delta}$ diodes (152). The inset shows transparent CuI thin films (*t* = ~120 nm) grown on glass. The $I_F/I_R$ value represents the rectification ratio. (*e*) Schematics of Ba$_{0.89}$K$_{0.11}$SnO$_3$/Ba$_{0.99}$La$_{0.01}$SnO$_3$ thin films. (*f*) Current-versus-voltage curves of Ba$_{0.89}$K$_{0.11}$SnO$_3$/Ba$_{0.99}$La$_{0.01}$SnO$_3$ diodes. Panels *e* and *f* adapted with permission from Reference (128); copyright 2016, AIP. Abbreviations: RT, room temperature; x-tal, single crystal.

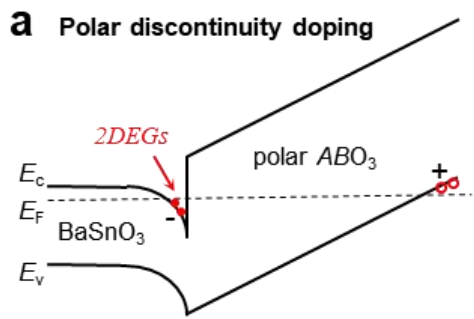
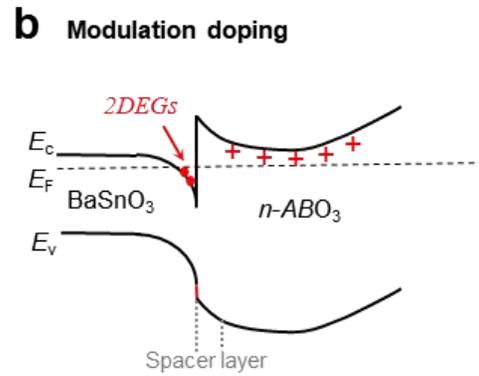
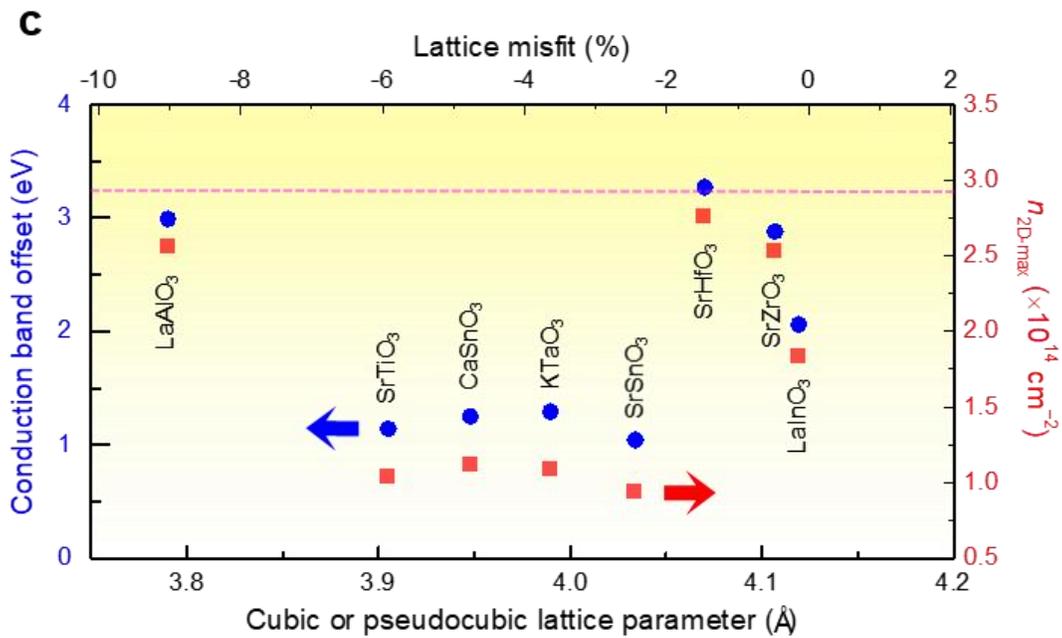
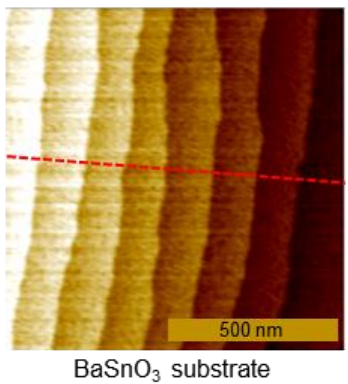
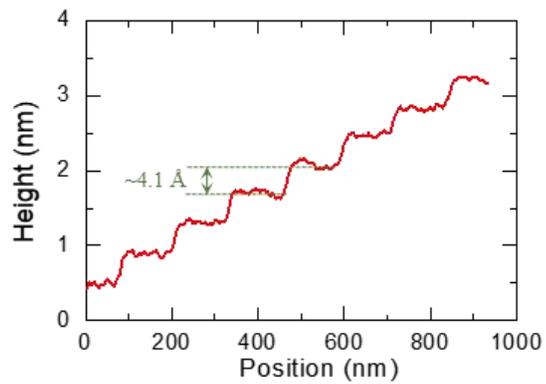

Figure 11 Proposed scenarios for realizing two-dimensional electron gases (2DEGs) at the heterointerface between a BaSnO$_3$ channel and other perovskite barrier materials. (*a*) A band diagram depicting polar discontinuity doping between a nonpolar BaSnO$_3$ and a polar *AB*O$_3$. (*b*) A band diagram describing modulation doping at a nonpolar heterointerface. (*c*) On the basis of the theoretical results of Schrödinger-Poisson simulations in Reference 97, the conduction band offsets (*blue circles*) and $n_{2D,max}$ (*red squares*) between BaSnO$_3$ and other *AB*O$_3$ barrier materials are replotted as a function of cubic or pseudocubic lattice parameters. The dashed magenta line indicates the $n_{2D,ideal}$, where all the electrons provided by polar discontinuity doping (0.5 electrons per in-plane unit cell) are confined in BaSnO$_3$. (*d*) An atomic force microscopy image showing the topography of the surface of the BaSnO$_3$(001) substrate (*left panel*) and the vertical profile taken along the dashed line (*right panel*). The surface exhibited atomically flat terraces and steps ~4.1 Å in height. Panel *c* adapted with permission from Reference 97; copyright 2016, AIP.

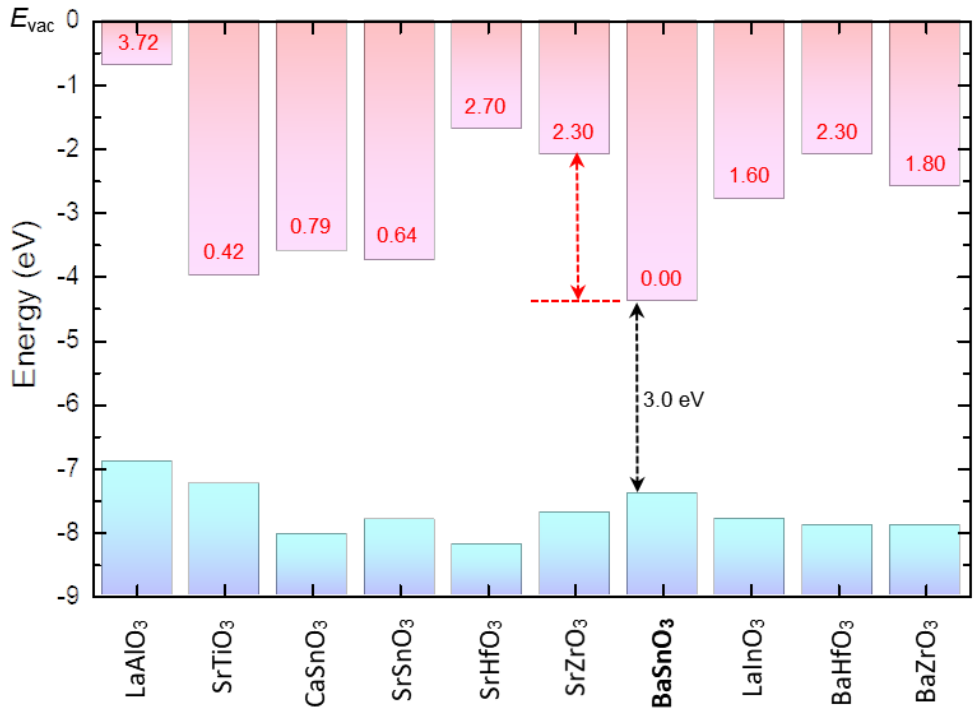

Figure 12 Summary of band alignments of various perovskite oxides, including BaSnO$_3$, as extracted from the available experimental data. The numbers reported in red are the conduction band offsets between BaSnO$_3$ and various perovskite oxides and reflect the position of the conduction band minimum.

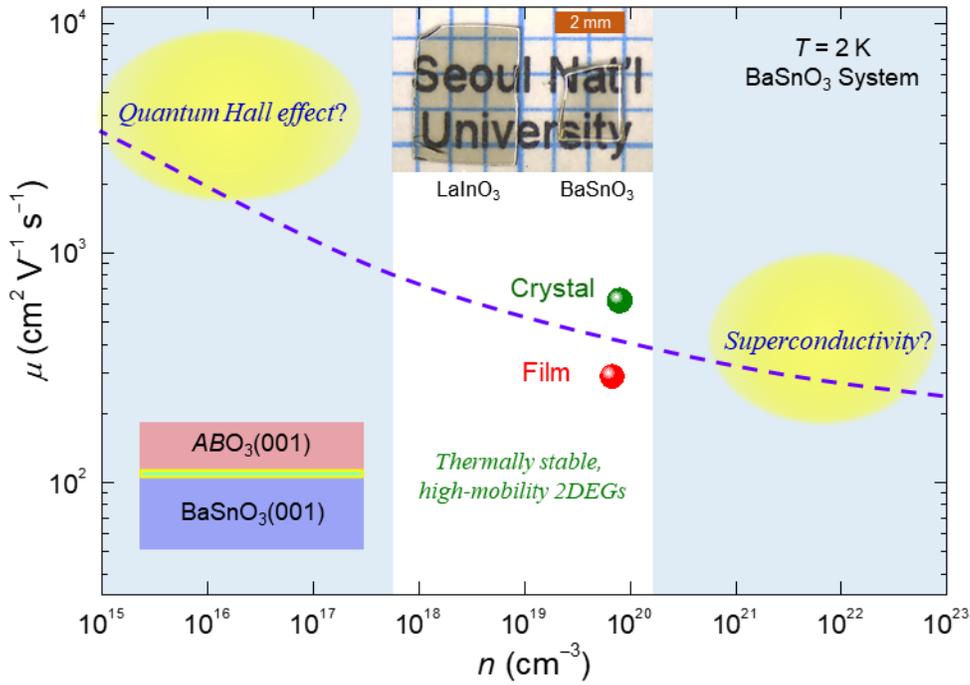

Figure 13 Expected electron mobility $\mu$ behavior and anticipated physical phenomena of the doped BaSnO$_3$ system at low temperatures. The dashed line represents the predicted $\mu$ from ionic dopant scattering; $\mu$ can increase more in a nondegenerate regime. The solid green and red circles indicate the best currently available experimental data for (Ba,La)SnO$_3$ single crystals and thin films, respectively. Two-dimensional electron gases (2DEGs) formed at the heterointerface between BaSnO$_3$ and other $ABO_3$ materials may exhibit fascinating quantum phenomena such as the quantum Hall effect and superconductivity in low- and high-doping regimes, respectively. To illustrate the importance of using the proper substrates, we show optical images of transparent LaInO$_3$ ($t$ = 300 μm) and BaSnO$_3$ ($t$ = 700 μm) crystals grown by our research group (*top middle*).

# Supplemental material

**Section S1.**

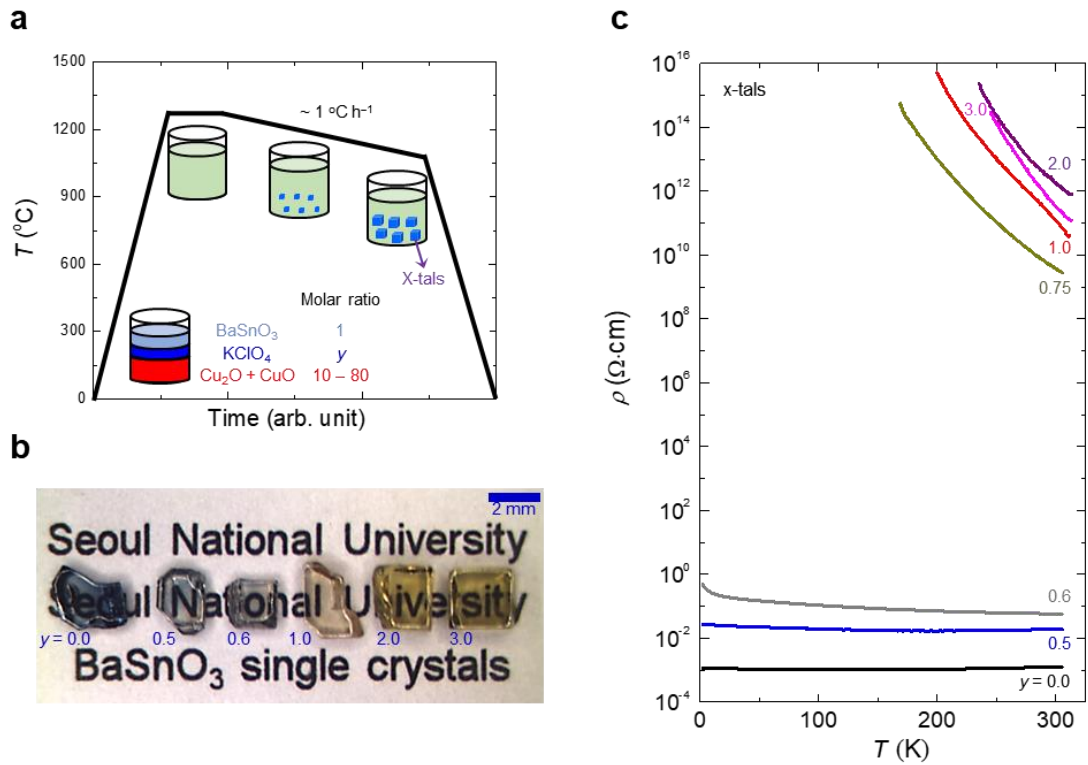

**Figure S1** The growth of insulating BaSnO$_3$ single crystals. (*a*) The applied temperature profile in the flux method with the molar ratio among the materials used. The mixture of BaSnO$_3$, Cu$_2$O+CuO, and KClO$_4$ with various molar ratio of BaSnO$_3$ : Cu$_2$O+CuO : KClO$_4$ = 1 : 10 – 80 : $y$ in the Pt crucible was fired for 5 hr at 1230 °C and then slowly cooled to 1070 °C by 1 °C h$^{-1}$ rate. It has been expected that KClO$_4$ can play a role of an oxidizer or a potassium ion (K$^+$) provider during the crystal growth (1-3) it can then provide additional oxygen to reduce oxygen vacancies or K$^+$ as acceptors to compensate the n-type carriers by oxygen vacancies in the BaSnO$_{3-\delta}$ crystals. (*b*) The photograph of the grown crystals with a different $y$ (0.0 – 3.0) in the flux. Although the crystals are mostly transparent, its color at the surface systematically changed from dark bluish to transparent yellowish color with the increase of the *y*. (*c*) Temperature-dependent electrical resistivity $\rho$ of the grown BaSnO$_3$ single crystals. The BaSnO$_3$ crystals with $y$ = 0.0, 0.5, and 0.6 exhibited a degenerately doped semiconducting (or metallic) behavior while those with $y$ = 0.75, 1.0, 2.0, and 3.0 exhibited an insulating behavior. Between $y$ = 0.6 and 0.75, the room temperature $\rho$ differs by $5 \times 10^{10}$, indicating there is a percolative transition.

## Section S2.

**Table S1** Comparison of cubic or pseudo-cubic lattice parameters, lattice misfit to BaSnO$_3$, and band gap of the cubic or orthorhombic perovskite that have been used or can be used potentially as a substrate for growing the BaSnO$_3$ films (4-27).

| | Cubic (C) or Orthorhombic | Orthorhombic | | | | Cubic | Lattitce misfit to BaSnO$_3$ (%) | Band gap (eV) | References in a main article |
| | | Lattice parameters (Å) | | | Pseudo-cubic lattice parameters (Å) | Cubic lattice parameters (Å) | | | |
| | | a | b | c | $a_{pc}$ | $a_c$ | | | |
|---|---|---|---|---|---|---|---|---|---|
| SrTiO$_3$ | Cubic | - | - | - | - | 3.905 | -5.13 | 3.26 | (16; 17) |
| CaSnO$_3$ | Orthorhombic | 5.5142 | 7.8816 | 5.6634 | 3.948 | - | -4.08 | 4.44 | (18; 19) |
| TbScO$_3$ | Orthorhombic | 5.466 | 7.917 | 5.731 | 3.958 | - | -3.84 | 6.10 | (20; 21) |
| KTaO$_3$ | Cubic | - | - | - | - | 3.990 | -3.07 | 3.64 | (22; 23) |
| SmScO$_3$ | Orthorhombic | 5.527 | 7.965 | 5.758 | 3.987 | - | -3.14 | 5.60 | (21; 24) |
| PrScO$_3$ | Orthorhombic | 5.608 | 8.025 | 5.780 | 4.021 | - | -2.30 | 5.70 | (21; 25) |
| SrSnO$_3$ | Orthorhombic | 5.709 | 8.065 | 5.703 | 4.034 | - | -1.99 | 4.00 | (18; 26) |
| LaScO$_3$ | Orthorhombic | 5.6803 | 8.0945 | 5.7907 | 4.053 | - | -1.54 | 5.80 | (27; 28) |
| SrHfO$_3$ | Orthorhombic | 5.752 | 8.134 | 5.765 | 4.070 | - | -1.11 | 6.50 | (29; 30) |
| PrInO$_3$ | Orthorhombic | 5.908 | 8.1538 | 5.6605 | 4.085 | - | -0.75 | 3.70 | D. H. Jang et al. unpublished |
| SrZrO$_3$ | Orthorhombic | 5.800 | 8.205 | 5.822 | 4.107 | - | -0.22 | 5.60 | (31; 32) |
| BaSnO$_3$ | Cubic | - | - | - | - | 4.116 | 0.00 | 3.00 | (1; 17) |
| LaInO$_3$ | Orthorhombic | 5.9414 | 8.2192 | 5.7249 | 4.119 | - | 0.08 | 5.00 | (149) |
| BaHfO$_3$ | Cubic | - | - | - | - | 4.1710 | 1.34 | 5.80 | (33; 34) |
| LaLuO$_3$ | Orthorhombic | 5.8259 | 8.3804 | 6.0218 | - | - | 1.77 | 5.50 | (35; 36) |
| BaZrO$_3$ | Cubic | - | - | - | - | 4.1930 | 1.87 | 5.30 | (37; 38) |

**Section S3.** Detailed explanations on various physical properties of doped $BaSnO_3$

To understand the gap nature and electronic structure of electron-doped $BaSnO_3$ system, we performed the optical experiments and density functional theory (DFT) calculations with local-density approximation (LDA) exchange-correlation functional. **Figure 7a** shows the band structures of $Ba_{0.963}La_{0.037}SnO_3$ and $BaSn_{0.963}Sb_{0.037}O_3$, indicating that the both La and Sb doping provide electronic states well-inside the conduction band of $BaSnO_3$ (21; 28). As the conduction band of $BaSnO_3$ is mainly composed of Sn 5$s$ states with the Sn-O antibonding character, the occupation in the antibonding state results in repulsive forces between Sn and O to lower the total energy of the crystal structure, thereby inducing expansion of lattice constant ($a_c$) upon carrier doping. The calculated electron effective mass ($m^*$) of the conduction band turns out to be ~$0.4m_0$ for both $Ba_{0.963}La_{0.037}SnO_3$ and $BaSn_{0.963}Sb_{0.037}O_3$. As the predicted $m^*$ from DFT calculations are rather sensitive to the approximation method (Section 5.1), its experimental determination becomes necessary.

**Figure 7b** shows the transmittance and optical absorption spectra for the $BaSnO_{3-\delta}$, $BaSn_{0.97}Sb_{0.03}O_3$, and $Ba_{0.98}La_{0.02}SnO_3$ single crystals. Transmittance of the $BaSnO_{3-\delta}$ crystal reached as high as 0.71 in a visible spectral region (1.8 – 3.1 eV) although its thickness $t = 68.7$ μm is rather high. The significant suppression of the transmittance as well as a sharp increase in the absorption coefficient ($\alpha$) near 3.1 eV reflects the development of the gap feature. The indirect and direct optical band gap ($E_g$) estimated by Tauc plot were 2.95 and 3.10 eV, respectively. DFT calculations by the hybrid functional seems almost consistent with those estimated $E_g$; Li et al. (29) and Liu et al. (30) reported indirect & direct band gaps as 2.82 & 3.17 eV and 2.6 & 3.0 eV, respectively. Transmittance of the $Ba_{0.98}La_{0.02}SnO_3$ ($t = 54$ μm) and $BaSn_{0.97}Sb_{0.03}O_3$ ($t = 52$ μm) crystals exhibits progressive suppression at a low frequency region, resulting in the Drude-type absorption tail due to the increase of free carriers. However, as the resultant $\alpha$ is less than 600 cm$^{-1}$ in the visible spectral region, the transmittance of $Ba_{0.98}La_{0.02}SnO_3$ and $BaSn_{0.97}Sb_{0.03}O_3$ with $t \leq 100$ nm is expected to be larger than ~0.8 (28). Moreover, the steep increase in $\alpha$ around 3 eV has slightly shifted to a higher energy with the increase of carrier doping. This is an evidence of the Burstein-Moss shift (31; 32), which is often observed in degenerate semiconductors. The difference in the $E_g$ ($\Delta E$) between undoped and doped cases is predicted as $\Delta E = h^2 (3n/\pi)^{2/3} / (8m^*)$, upon assuming a parabolic dispersion near the conduction band minimum. Here, $h$ is Plank constant and $n$ is the carrier density. The estimated $m^*$ was 0.61 $m_0$ according to our former study (21; 28). This value was somewhat bigger than the theoretically estimated value of ~0.4$m_0$ from our LDA calculations and other values of 0.06 – 0.2$m_0$ (Section 5.1). In subsequent experimental investigation of the (Ba,La)SnO$_3$ thin films, both the Burstein-Moss shift and the Drude weight calculation consistently produced $m^*$ = ~0.35 $m_0$ (33), suggesting that the experimental $m^*$ values could be sensitive to the sample kinds or other physical origins. Related to this, hard X-ray photoemission spectroscopy (XPS) study by Lebens-Higgins et al. recently reported the presence of significant non-parabolic band effect and band renormalization in a high doping regime (34), implying that the estimated $m^*$ by the Burstein-Moss shift in a parabolic approximation can be apparently bigger than the actual value.

Another interesting physical quantity of BaSnO$_3$ as a potential platform for developing perovskite based-oxide electronics is the thermal conductivity ($\kappa$). For applications as a substrate material, a larger $\kappa$ could be helpful to reduce heating of devices and to guarantee reliable operations. **Figure 7c** shows the temperature-dependent $\kappa$ of the BaSnO$_{3-\delta}$ single crystal with $n$ = ~$10^{18}$ cm$^{-3}$ based on the $3\omega$ method (35). $\kappa$ of the BaSnO$_{3-\delta}$ crystal at 300 K is 0.132 W cm$^{-1}$ K$^{-1}$. As the temperature is lowered, the $\kappa$ approaches 1.44 W cm$^{-1}$ K$^{-1}$ at 22 K, being 10 times larger than the value at 300 K. The ten times increase of $\kappa$ means that the BaSnO$_{3-\delta}$ crystal has high crystallinity, which renders the phonons as a main source of the heat transport. This is also verified by the $\kappa$ behavior of being proportional to $T^{-1}$, while an electronic contribution was estimated to be much smaller than 4 % of the total $\kappa$. In comparison, $\kappa$ of the BaSnO$_{3-\delta}$ crystal was overall larger than those of other perovskite oxides, e.g., SrTiO$_3$ (36), KTaO$_3$ (37), and LaMnO$_3$ (38) crystals as compared in **Figure 7c**.

We find that carrier doping in BaSnO$_3$ produces significant local lattice distortion as evidenced by Raman spectroscopy. **Figure 7d** shows the Raman spectra of metallic Ba$_{0.99}$La$_{0.01}$SnO$_3$, BaSnO$_{3-\delta}$, and highly insulating BaSnO$_3$ singe crystals. All three single crystals turned out to show very similar X-ray scattering profiles. We verified from the refinement that all the scattering data could be well fitted by the cubic perovskite structure. In an ideal cubic perovskite, the first order Raman modes should be absent by symmetry. Consistently, we found almost no Raman active mode in the highly insulating BaSnO$_3$ single crystal except a small intensity at 833 cm$^{-1}$, which might be a local vibrational mode related to intrinsic defects, e.g., cation vacancies. In sharp contrast, both metallic Ba$_{0.99}$La$_{0.01}$SnO$_3$ and BaSnO$_{3-\delta}$ single crystals with $n = 1.0 \times 10^{20}$ and $4.0 \times 10^{18}$ cm$^{-3}$, respectively, exhibited quite strong Raman modes at ~138 cm$^{-1}$, ~250 – 350 cm$^{-1}$, and ~615 – 650 cm$^{-1}$. Moreover, both of the Raman spectra are quite similar each other. This result implies the presence of electron carriers in the BaSnO$_3$ system produces local lattice distortion, which possibly breaks the cubic symmetry locally. One possible scenario is that occupation of the antibonding state via electron doping can induce the expansion of the Sn-O bond length, presumably involving the breathing mode distortion. This expansion in the bond length will likely induce reduction of the Sn-O-Sn bonding angle from an ideal 180 °, which then results in local symmetry lowering toward the orthorhombic structure. The main modes observed in the Raman spectra in **Figure 7d** are likely to be stretching (~615 – 650 cm$^{-1}$), bending (~250 – 350 cm$^{-1}$), and external (~138 cm$^{-1}$) phonon modes expected of SnO$_6$ octahedra in the lower crystal symmetry than cubic structure, i.e., the orthorhombic or tetragonal symmetry (39; 40). Our results naturally imply that electron-phonon coupling in doped-BaSnO$_3$ can be significantly large.

**Figure 7e** shows the temperature-dependent $n$ (electron carrier density), $\mu$ (electron mobility), and $\rho$ (electrical resistivity) to understand the electrical transport properties of Ba$_{0.995}$La$_{0.005}$SnO$_3$, BaSn$_{0.97}$Sb$_{0.03}$O$_3$, and BaSnO$_{3-\delta}$ single crystals as well as Ba$_{0.99}$La$_{0.01}$SnO$_3$/SrTiO$_3$(001) and Ba$_{0.995}$La$_{0.005}$SnO$_3$/BaSnO$_3$(001) thin films (3; 21; 28). First, all the $n$ is almost temperature independent in those crystals and films. $\rho$ in both crystals and films mostly increases with increase of temperature; this typical metallic behavior clearly supports that they remain in a degenerately doped semiconducting regime. $\rho$ of the Ba$_{0.995}$La$_{0.005}$SnO$_3$ crystal decreases nearly by a factor of two upon being cooled from room temperature to 2 K due to the increase of $\mu$ by the same factor. The $\mu$ increase at low temperatures mainly originates from the reduction of the acoustic phonon scattering, indicating that the scattering rate due to acoustic phonons is approximately half of the total scattering rate at room temperature. Moreover, the resistivity (residual-resistivity ratio) of the BLSO films on the $SrTiO_3$(001) and $BaSnO_3$(001) substrates are generally larger (smaller) than that of $Ba_{0.99}La_{0.01}SnO_3$ crystals, implying that there exist extra-scattering sources in the BLSO/$SrTiO_3$(001) and BLSO/$BaSnO_3$(001) films. We attribute the extra-scattering sources to cation vacancies or cation site mixing in the BLSO/$BaSnO_3$(001) films (3) and to a large fraction of threading dislocations or grain boundaries in the BLSO/$SrTiO_3$(001) films (21). We discuss the major scattering sources of the BLSO films in more detail in Section 3 in the main review.

**Section S4.** Detailed explanations on transistors based on $BaSnO_{3-\delta}$

**Figures 9a,b** show a schematic drawing and a top-view image of the $BaSnO_{3-\delta}$-TFT under a polarization microscope, respectively. Our approach still poses a technical difficulty that the substrate is too small to use a stencil mask for defining the channel area. The device was fabricated on the insulating $BaSnO_3$(001) substrate and used the epitaxial $BaSnO_{3-\delta}$ film ($t$ = 90 nm) as a channel layer. Because a stencil mask could not be used to define the smaller channel area, the channel layer covered all the substrate area (~1 × 1 mm$^2$). Amorphous $Al_2O_3$ (50 nm), Au/Ti/ITO (20/5/20 nm), and Au (50 nm) films were used as a gate oxide, source & drain, and gate electrode, respectively. **Figures 9e,f** show the output (drain-to-source current $I_{DS}$ vs. drain-to-source voltage $V_{DS}$) and transfer ($I_{DS}$ vs. gate-to-source voltage $V_{GS}$) characteristics of the $BaSnO_{3-\delta}$-TFT, respectively. From the output characteristics, the channel was n-type and carriers were generated by a positive $V_{GS}$. $BaSnO_{3-\delta}$-TFT exhibited ohmic properties at a low $V_{DS}$ regime and current saturation at a high $V_{DS}$ regime. The threshold voltage $V_T$ was 1.52 V as extracted from a linear fit in the $I_{DS}^{0.5}$ vs. $V_{GS}$ plot, indicating that the $BaSnO_{3-\delta}$-TFT operates in an enhancement mode: the positive value of $V_T$ supports that the Fermi level $E_F$ in the $BaSnO_{3-\delta}$ channel is located below the energy level of the conduction band minimum.

The transfer characteristic shows that the off-current was on the order of 100 nA, and the on-off current ratio and subthreshold swing ($S$) were ~3 × 10$^3$ and 1.67 V dec$^{-1}$, respectively. The rather low on-off ratio likely stems from the large off-current induced by the large channel area in this particular TFT. In the off-state, the conductivity of the BaSnO$_{3-\delta}$ channel was found to be ~4.0 × 10$^{-3}$ S cm$^{-1}$, which represents $n$ = ~2.6 × 10$^{14}$ – 2.6 × 10$^{15}$ cm$^{-3}$ upon assuming $\mu$ = ~100 – 10 cm$^2$ V$^{-1}$s$^{-1}$. This implies that a single-crystalline BaSnO$_{3-\delta}$ film with $n$ < 10$^{17}$ cm$^{-3}$ is fabricated. The calculated $\mu_{FE}$ in the BaSnO$_{3-\delta}$-TFT from the transfer characteristics was as high as ~45.0 cm$^2$ V$^{-1}$s$^{-1}$. Although the device has been fabricated rather crudely even without defining the small channel area, the obtained $\mu_{FE}$ is quite high; it is much larger than the typical values (< 1 cm$^2$/V·s) reported in other TFTs based on the perovskite materials (41; 42). Furthermore, it is already comparable to those of the well-known TFTs based on the transparent binary oxides; for example, $\mu_{FE}$ was ~40 cm$^2$ V$^{-1}$s$^{-1}$ in the ZnO-based TFT (43), in which a similar top-gate TFT was fabricated with an amorphous Al$_2$O$_3$ gate insulator. The large $\mu_{FE}$ in the BaSnO$_{3-\delta}$-TFT is mainly attributed to the reduced dislocation densities in the channel grown on the BaSnO$_3$(001) substrate. For comparison, we also fabricated the almost same BaSnO$_{3-\delta}$-TFT on the STO(001) substrate without buffer layer to find $\mu_{FE}$ = ~5 cm$^2$ V$^{-1}$s$^{-1}$ only.

In order to reduce the off-current coming from the large channel area in the initial devices, we also tried to adopt the Si stencil-masks to better define the BaSnO$_{3-\delta}$ (60 nm) channel area. **Figures 9c,d** show a schematic drawing and a top-view image of the BaSnO$_{3-\delta}$-TFT fabricated by the Si stencil-masks. The channel length and the width of the device were 50 and 200 μm, respectively. **Figure 9g** summarizes a transfer characteristic with $\mu_{FE}$ of the modified BaSnO$_{3-\delta}$-TFT. In this case, the on-off current ratio was increased up to 1.2 × 10$^6$ as expected from the well-defined channel area while the resultant $\mu_{FE}$ and the $S$ were found to be 48.7 cm$^2$ V$^{-1}$s$^{-1}$ and 1.55 V dec$^{-1}$, respectively, which were somewhat similar to the cases in **Figure 9f**. On the other hand, this device operated in the depletion mode with a $V_T$ = −22.2 V. The curvatures of $I_{DS}$ found at the $V_{GS}$ = −6 and 16 V indicate that impurity states of a shallow donor-type have been formed in the BaSnO$_{3-\delta}$ channel layer. The shallow donors can be formed by inherent small $V_O$ generated in the PLD film-growth or hydrogen contamination present during the deposition of amorphous Al$_2$O$_3$ gate oxides by atomic layer deposition. Therefore, it is still expected that the performance of TFTs based on the insulating BaSnO$_3$(001) substrate could be improved further with fine control of defect levels or with use of different gate oxides and so on.

Another direction in the researches of TFTs is to use a buffer layer on $SrTiO_3$(001) substrate before depositing a channel layer to reduce threading dislocations. In 2014, Park et al. reported a use of $BaSnO_3$ buffer layers (110 nm) to fabricate TFT composed of a La-doped $BaSnO_3$ channel (0.14 and 0.5 %) and an amorphous $Al_2O_3$ gate oxide (44). The $\mu_{FE}$, on-off current ratio, and $S$ value were 17.8 $cm^2 V^{-1} s^{-1}$, $10^5$, and 3.2 V $dec^{-1}$, respectively. The same group also reported a similar TFT structure except using a $HfO_2$ gate oxide to find $\mu_{FE}$, on-off current ratio, and $S$ value as 24.9 $cm^2 V^{-1} s^{-1}$, $6 \times 10^6$, and 0.42 V $dec^{-1}$, respectively (45). Although there was an improvement of $\mu_{FE}$ value by using a gate oxide with higher dielectric constant ($\varepsilon_{r,HfO2}$ = 23.7 and $\varepsilon_{r,Al2O3}$ = 8.0), $\mu_{FE}$ values in these cases were still smaller than our TFT devices using the $BaSnO_3$(001) substrate. This indicates that the dislocation scattering could still reduce the $\mu_{FE}$ albeit using the buffer layer. Fujiwara et al. also reported $BaSnO_3$-TFT with $BaSnO_3$ (141 nm) and $Sr_{0.5}Ba_{0.5}SnO_3$ buffer layers (14 nm) on the $SrTiO_3$(001) substrate (46). In particular, they found that additional deposition of a $Sr_{0.5}Ba_{0.5}SnO_3$ layer (10 nm) on top of a $BaSnO_3$ channel layer could result in enhanced $\mu_{FE}$ = 52 $cm^2 V^{-1} s^{-1}$. The authors proposed that the $Sr_{0.5}Ba_{0.5}SnO_3$ layer played a role of confining accumulated carriers at the heterointerface between $BaSnO_3$ and $Sr_{0.5}Ba_{0.5}SnO_3$ as it has a higher band gap by ~0.4 eV than in $BaSnO_3$. This work demonstrates that the FET characteristics can be improved by such heterointerface engineering. Kim et al. in 2015 fabricated the heterojunction TFT employing all perovskite oxides, i.e., a BLSO channel (La = 0.07 %) and a $LaInO_3$ gate oxide grown on the $BaSnO_3$ buffer (100 nm)/$SrTiO_3$(001) substrate (47). This device achieved a large on-off ratio = ~$10^7$ along with $\mu_{FE}$ = 90 $cm^2 V^{-1} s^{-1}$. The use of a $LaInO_3$ film with $\varepsilon_r$ = 38.7 and rather a large $E_g$ (~5.0 eV) seems to be also useful in accumulating larger carriers in the interface of the channel. Based on all the progresses, it is expected that the new $BaSnO_3$-based TFT with high $\mu_{FE}$ could offer a solution for next-generation high-speed, multi-functional transparent devices.

**Section S5.** Detailed explanations on pn junctions based on doped $BaSnO_3$

Because the thin films grown on the $SrTiO_3$(001) substrates are strongly subject to the dislocation-limited scattering, we first studied basic properties of the polycrystalline $BaSn_{0.9}Co_{0.1}O_3$ target; the activation energy $E_a$ was found to be 0.32 eV from the Arrhenius plot of the electrical resistivity $\rho$. By assuming p-type carrier density $p = p_0\exp(-E_a/k_BT)$ with $p_o = 1.44 \times 10^{21}$ cm$^{-3}$ (nominal dopants), $p$ at room temperature were estimated as $7.4 \times 10^{15}$ cm$^{-3}$. From $\rho$ of 500 Ω·cm at 300 K, the hole mobility is then estimated to be 1.7 cm$^2$ V$^{-1}$·s$^{-1}$.

**Figures 10a** shows schematic drawing of pn junctions of an epitaxial p-$BaSn_{0.9}Co_{0.1}O_3$ film grown on the n-type single crystal of $BaSn_{0.97}Sb_{0.03}O_3$ ($n = \sim 1.2 \times 10^{20}$ cm$^{-3}$) with a rather large lateral junction area of ~1 mm$^2$. The $BaSn_{0.9}Co_{0.1}O_3$ film was grown by the PLD method at 790 °C and oxygen partial pressure of 0.1 Torr. The grown film was confirmed to show the p-type character as $\rho$ under $O_2$ (Ar) atmosphere decreased (increased) at high temperature. **Figure 10b** shows the *j-V* characteristic of p-$BaSn_{0.9}Co_{0.1}O_3$/n-$BaSn_{0.97}Sb_{0.03}O_3$ diode showing the rectification ratio of $2.9 \times 10^3$ (±2 V) at room temperature. We found that pn diode showed a similar rectification ratio even after the thermal annealing of 200 °C. This results demonstrates that it is straightforward to fabricate thermally stable pn diodes based on the $BaSnO_3$ system.

Kim et al. also reported a similar pn-junction based on the Ba-site doped, p-$(Ba_{0.89}K_{0.11})SnO_3$ film grown on n-$Ba_{0.99}La_{0.01}SnO_3$ (100nm), which were grown on $BaSnO_3$ buffer (5nm)/$SrTiO_3$(001) (48). They reported that at room temperature, K$^+$ doping resulted in $p = \sim 1.0 \times 10^{13}$ cm$^{-3}$ and the rectification ratio of 1135 (±2 V) was found in a pn diode with junction size of ~0.5 mm$^2$. Moreover, they found that the pn diode became thermally stable up to 300 °C, and the rectification reached 1140 with the more activated p-type carriers. All these results along with those in **Figures 10e,f** show that thermally stable pn-junctions can be fabricated from p-type TOSs based on the $BaSnO_3$ system (either by Ba or Sn site doping), increasing the potential of the perovskite stannate system for applications toward optoelectronic devices. Yet, both of above results have not shown any generation of light in the pn junctions under the forward bias. This implies that the indirect gap nature in the $BaSnO_3$ system is not efficient to have strong carrier recombination process. Therefore, to tune the material properties toward having more direct $E_g$ nature should be a necessary step for future optoelectronic applications.

Another interesting question related to, e.g., solar cell applications is whether transparent pn-junction can be fabricated. CuI has been recently known to be a promising transparent p-type material with $p = \sim 4.3 \times 10^{16}$ cm$^{-3}$ and a direct $E_g$ of ~3.1 eV (49). Inspired by the successful transparent pn junction fabrication in combination with n-ZnO film (50), we fabricated p-CuI (1000 nm)/n-BaSnO$_{3-\delta}$ (100 nm) films on a Ba$_{0.96}$La$_{0.04}$SnO$_3$ buffer (100 nm)/SrTiO$_3$(001) substrate with a lateral junction size of 0.04 mm$^2$ (**Fig. 10c**) (Lee JH et al. under revision). For the growth of CuI, a Cu thin film on the SrTiO$_3$(001) substrate was firstly deposited by sputtering method and was then reacted with I$_2$ in a sealed quartz at 80 °C for 30 min. An X-ray $\theta$-$2\theta$ scan revealed that the CuI film has polycrystalline nature with a cubic zinc-blende structure phase ($\gamma$-CuI). **Figure 10d** shows the *j-V* characteristic of p-CuI/n-BaSnO$_{3-\delta}$ diode showing the rectification ratio of $8.6 \times 10^4$ at room temperature. It is noteworthy that the p-CuI/n-BaSnO$_{3-\delta}$ diode has exhibited the increase of both rectification ratio and saturation current over those of pn junctions based on the p-type BaSnO$_3$ (p-BaSn$_{0.9}$Co$_{0.1}$O$_3$ and p-Ba$_{0.89}$K$_{0.11}$SnO$_3$). The increase of p–type carrier density seems to be a main origin for the increase of the rectification ratio. Upon further reducing the thickness of CuI, it is likely that the pn diode based on p-CuI/n-BaSnO$_{3-\delta}$ heterostructures with a good interface property can be fabricated. We also found that the junction properties are nearly same for more than 180 days after repeated *I-V* measurements of the junction.

**Section S6.** Detailed explanations on solar cells and photoconductors based on BaSnO$_{3-\delta}$

With its superior $\mu$, the BaSnO$_3$ system might find important applications in the photovoltaic cells. Researchers have indeed fabricated BaSnO$_3$ based nanoparticles to use in dye-sensitized solar cells (DSSCs) (51-53) and perovskite solar cells (54; 55). With its simple processing, high price/performance ratio and a wide range of applications, DSSCs are rapidly expanding the territory as emerging solar cells (56-58). The properties of DSSCs have depended strongly on a mesoporous layer containing nanoparticulate oxides with a wide-band gap, e.g., TiO$_2$, to amplify greatly the surface area and to transport easily electron carriers (59). Besides, the perovskite solar cells, in which organic perovskite halides are used instead of the dye materials, are currently drawing much attention because they have great potential of achieving higher efficiencies and the very low production costs, becoming commercially attractive (60-62). In the perovskite solar cells, the $TiO_2$ particles have been often used to transfer electron carriers easily (63).

In 2010, Li et al. reported the use of porous $BaSnO_3$ nanoparticles alternative to $TiO_2$ nanoparticles in the DSSCs to find the energy conversion efficiency of 0.12 % (51). In 2013, Shin et al. and Kim et al. reported efficiencies of 6.2 % and 5.2 %, respectively, in the DSSCs using the $BaSnO_3$ nanoparticles (52; 53). In 2016, Zhu et al. similarly substituted mesoporous $BaSnO_3$ for the mesoporous $TiO_2$ to use as the electron transport agent in the emergent perovskite solar cells (54), finding the increased power-conversion efficiency of 12.3 % ($BaSnO_3$) from 11.1 % ($TiO_2$). The main origin for the increase was attributed to the high mobility of $BaSnO_3$. Most recently in 2017, photostable perovskite solar cells with power-conversion efficiency of 21.2 % was reported by use of BLSO transporting layer, which was prepared by heat treatments under 300 °C (55). Thus, the nanoparticles of $BaSnO_3$ used in the mesoporous layer of DSSCs or perovskite solar cells could be a competitive candidate in future upon further optimizing the fabrication procedure.

With its superior mobility and wide $E_g$ of $BaSnO_3$ system, a photoconductive sensor at UV and deep visible spectrum region is another direction of potential application of the $BaSnO_3$ system. In 2016, Park et al. made the first attempt toward this to compare the photoconductivity of $BaSnO_3$ film on MgO(001) substrate ($BaSnO_3$/MgO(001)) with that of $SrTiO_3$ film on MgO(001) substrate ($SrTiO_3$/MgO(001)) (64). Although the photoconductivity was higher in $BaSnO_3$/MgO(001) possibly due to the higher mobility, $BaSnO_3$/MgO(001) showed strong persistent conductivity for many hours after removal of light exposure whereas $SrTiO_3$/MgO(001) showed little persistent photoconductivity. The persistent photoconductivity is an obvious drawback for a potential photoconductive sensor, for which the defect states in the film is mainly responsible. Therefore, removal or reduction of defects in the $BaSnO_3$ film by use of, e.g., a proper substrate seems to be another challenge to overcome in the researches toward realizing a photoconductor sensor.

**Section S7.** How to determine band alignment in Figure 12.

In order to summarize experimental band alignment of BaSnO$_3$ between candidate materials for a reference for future researches, we first note that the known electron affinity of SrTiO$_3$ (4.0 eV) (65) was used to determine its conduction band minimum with respect to the vacuum level. Chambers et al. indeed determined accurately the valence band offsets (VBO) of SrTiO$_3$ and LaAlO$_3$ with respect to BaSnO$_3$ (5). They then estimated the CBOs of SrTiO$_3$ and LaAlO$_3$ with respect to BaSnO$_3$ in combination with the known $E_g$ of SrTiO$_3$ and LaAlO$_3$. In **Figure 12**, we used the known information as it is to plot the band alignments of BaSnO$_3$ between SrTiO$_3$ and LaAlO$_3$.

To determine the band alignment of BaSnO$_3$ between SrZrO$_3$, the VBO between SrTiO$_3$ and SrZrO$_3$ as measured by XPS and the known $E_g$ (5.6 eV) of SrZrO$_3$ (20) was used. For the cases of SrSnO$_3$ and CaSnO$_3$, we first noticed that Dorenbos et al. provided the band alignments of SrSnO$_3$, CaSnO$_3$, and BaSnO$_3$ as estimated by the chemical shift model and the experimental reflection, luminescence and absorption spectra (66). However, the proposed $E_g$ of the three compounds were overestimating the more accurate experimental values determined by optical transmission measurements by 0.5 – 0.75 eV. Therefore, we only adopted the VBOs and CBOs of SrSnO$_3$ and CaSnO$_3$ with respect to BaSnO$_3$ from the results by Dorenbos et al. to draw the band alignments of SrSnO$_3$ and CaSnO$_3$ in **Figure 12**. For the band alignment of LaInO$_3$, we used recent reports by Kim et al, in which the CBO of LaInO$_3$ was determined by the estimation of a tunneling barrier in the *I-V* measurements (67). The VBO of LaInO$_3$ was determined by the $E_g$ of 5.0 eV as reported by Kim et al. It should be reiterated that the reported $E_g = 5.0$ eV in the LaInO$_3$ film is clearly different from that of the single crystal ($E_g = 4.3$ eV; **Figure *3a***), presumably due to the strain effect from the substrate. Therefore, the $E_g$ and/or dielectric properties of a LaInO$_3$ thin film should be investigated for each substrate material employed in the film growth for, e.g., accurate determination of $n_{2D}$ in **Figure 11*c***.

For drawing band alignments of BaSnO$_3$ between SrHfO$_3$, BaHfO$_3$, and BaZrO$_3$, we first used the experimental fact that SrTiO$_3$ has almost the same electron affinity with Si (68). The VBO between SrHfO$_3$ and Si was indeed determined by the photoelectron spectroscopy (18). We used the VBO of SrHfO$_3$ with respect to Si and its known $E_g$ to draw the band alignment of SrHfO$_3$ in **Figure 12**. Similarly, the VBOs among SrHfO$_3$, BaHfO$_3$, and BaZrO$_3$ were determined by XPS and X-ray absorption spectroscopy (23). We used the VBOs of the three materials and the $E_g$ to draw the band alignments of BaHfO$_3$, and BaZrO$_3$ (23; 27)